\documentclass{aa}  

\usepackage{graphicx}
\usepackage{txfonts}
\usepackage{lscape}
\usepackage[]{hyperref}
\usepackage{color}
\hypersetup{breaklinks=true,colorlinks=true,urlcolor=blue, linkcolor=blue, citecolor=blue,bookmarksopen=true}

\newcommand{\roph}{$\rho$-Oph}

\newcommand{\rbra}{Br$_\alpha$/Br$_\gamma$}
\newcommand{\rpfg}{Pf$_\gamma$/Br$_\gamma$}
\newcommand{\RV}{R$_{\mathrm V}$}
\newcommand{\AV}{A$_{\mathrm V}$}
         
\newcommand{\AK}{A$_{\mathrm K}$} 
\newcommand{\AKlocal}{A$_{\mathrm K,local}$}

\newcommand{\ABra}{A$_{4.05\mu}$}
\newcommand{\ABrg}{A$_{2.17\mu}$}

\newcommand{\Lsun}{L$_{\odot}$}
\newcommand{\Lbol}{L$_{\mathrm{bol}}$}
\newcommand{\Tbol}{T$_{\mathrm{bol}}$}

\newcommand{\Lstar}{L$_{\star}$}

\newcommand{\Bra}{Br$_\alpha$}
\newcommand{\Brg}{Br$_\gamma$}

\newcommand{\Pfg}{Pf$_\gamma$}
\newcommand{\LBra}{L(Br$_\alpha$)}
\newcommand{\LBrg}{L(Br$_\gamma$)}

\newcommand{\LPfg}{L(Pf$_\gamma$)}

\newcommand{\Lacc}{L$_{\mathrm {acc}}$}
\newcommand{\Lline}{L$_{{\mathrm line}}$}

\newcommand{\simless}{\mathbin{\lower 3 pt\hbox {$\rlap{\raise 5pt\hbox{$\char'074$}}\mathchar"7218$}}} 
\newcommand{\simgreat}{\mathbin{\lower 3pt\hbox {$\rlap{\raise 5pt\hbox{$\char'076$}}\mathchar"7218$}}} 

\begin{document}

   \title{The accretion luminosity of Class~I protostars}

   \author{L. Testi
          \inst{1, 2}
          \and
          A. Natta\inst{3}
          \and
          S. Gozzi\inst{1}
          \and
          C.F. Manara\inst{4}
          \and
          J.P. Williams\inst{5}
          \and
          R. Claes\inst{4}
          \and
          U. Lebreuilly\inst{6}
          \and
          P. Hennebelle\inst{6}
          \and
          R. Klessen\inst{7,8,9,10}
          \and
          S. Molinari\inst{11}
          }

   \institute{
             Alma Mater Studiorum -- Università di Bologna, Dipartimento di Fisica e Astronomia ``Augusto Righi'', Via Gobetti 93/2, I-40129, Bologna,
             Italy
             \and
             INAF-Osservatorio Astrofisico di Arcetri, Largo E. Fermi 5, I-50125 Firenze, Italy
             \and
             Astronomy \& Astrophysics Section, School of Cosmic Physics, Dublin Institute for Advanced Studies, 31 Fitzwilliam Place, Dublin D02 XF86, Ireland
             \and
             European Southern Observatory, Karl-Schwarzschild-Strasse 2, D-85748 Garching bei München, Germany
             \and
             Institute for Astronomy, University of Hawaii, Honolulu, HI 96822, USA
             \and Universit\'e Paris-Saclay, Universit\'e Paris-Cit\'e, CEA, CNRS, AIM, 91191 Gif-sur-Yvette, France
             \and Universit\"{a}t Heidelberg, Zentrum f\"{u}r Astronomie, Institut f\"{u}r Theoretische Astrophysik, Albert-Ueberle-Str. 2, 69120 Heidelberg, Germany
             \and
             Universit\"{a}t Heidelberg, Interdisziplin\"{a}res Zentrum f\"{u}r Wissenschaftliches Rechnen, Im Neuenheimer Feld 225, 69120 Heidelberg, Germany
             \and
             Harvard-Smithsonian Center for Astrophysics, 60 Garden Street, Cambridge, MA 02138, USA
             \and
             Elizabeth S. and Richard M. Cashin Fellow at the Radcliffe Institute for Advanced Studies at Harvard University, 10 Garden Street, Cambridge, MA 02138, USA
             \and 
             INAF - Istituto di Astrofisica e Planetologia Spaziali, Via Fosso del Cavaliere 100, I-00133 Roma, Italy
             }

   \date{\today}

 \abstract
 {The value of the accretion luminosity during the early phases of star formation is crucial information that aids in understanding how stars form, but it is still very difficult to obtain.}
 {We have developed a new methodology to measure accretion luminosity using mid-infrared hydrogen recombination lines and applied it to a limited sample of Class~I protostars in the Taurus and Ophiuchus star-forming regions.}
 {We adopted the commonly used assumption that the properties of disk-protostar accretion in Class~I objects is similar to the disk-star accretion in Class~II objects. Using simultaneous observations of three hydrogen recombination lines, \Brg, \Pfg, and \Bra, we derived the mean intrinsic line ratios, and we verified that these are constant across the probed range of photospheric and accretion properties. We established correlations between the line luminosities and accretion luminosity.}
 {We measured the extinction toward the line emission regions in Class~I protostars, comparing the observed line ratios to the Class~II mean values. We then derived the Class~I accretion luminosities from the established Class~II correlations. We find that the accretion luminosity dominates the bolometric luminosity for the more embedded protostars, corresponding to lower values of the bolometric temperature. As the bolometric temperature increases above $\sim$700~K, there is a sharp drop of the contribution of the accretion from the bolometric luminosity.}
 {Our findings are in qualitative agreement with numerical simulations of star formation. We suggest our methodology be applied to larger and more statistically significant samples of Class~I objects to obtain a more detailed comparison. Our results also suggest that by combining multiple infrared line ratios, it will be possible to derive a more detailed description of the dust extinction law in protostellar envelopes.} 


   \maketitle
%

\section{Introduction}
\label{sec:Intro}

It is widely accepted that planets form in protoplanetary disks, which are the byproduct of the star formation process \citep{1987ARAA..25...23S}. The study of disks around pre-main sequence stars \citep[{so-called Class II disks; for the original definition of the young stellar objects classification scheme, see}][]{1987IAUS..115....1L,1993ApJ...406..122A} has shown that in this phase, the disks are slowly evolving relics of more massive and dynamical structures that led to the rapid assembly of the central star and the formation of the planetary cores \citep[e.g.,][and references therein]{Testi2022,2023ASPC..534..539M}. Though the Class II disks are important for understanding disk-planet interactions and the dissipation of disks, it is in the early phases (Class 0 and I) that planet formation most likely begins and most of the action takes place. Recent numerical simulations of star formation that include non-ideal magneto-hydrodynamics (MHD) and feedback effects and resolve the disk formation phase and extend up to the ages of the youngest Class~I have shown that young star-disk systems are expected to have very high accretion luminosities that completely dominate the radiative output from the central object as well as the thermal structure of the young disks \citep{2021ApJ...922...36L,2021MNRAS.501.5873W,2024A&A...683A..13L,2025A&A...696A.238A}. Observationally, the disk-(proto)star interaction processes are still very poorly known in young embedded disks, yet they may play a critical role in shaping the overall initial conditions for planet formation \citep[e.g.,][]{2020A&A...635A..67H} and for the subsequent evolution of the star \citep[e.g.,][]{2010A&A...521A..44B}. The lack of understanding is exemplified by the well-known ``luminosity deficit” of protostars that  seem to be under-luminous compared to the theoretically predicted accretion rates necessary to assemble the central protostellar mass within the duration of the embedded evolutionary phase \citep[e.g.,][]{1990ApJ...349..197K, 2009ApJS..181..321E}.
{Several solutions for this conundrum have been proposed over the years, including a large spread of the accretion rates, both in time and environment conditions, leading to variable timescales to assemble the final stars \citep{2023ASPC..534..355F}. These effects result in a large spread of accretion luminosities for protostars, as observed. As noted by \citet{2023ASPC..534..355F}, the observational characterization of the accretion luminosity in these early stages is still limited. In this context, it is essential to understand the contribution of accretion luminosities to the total bolometric luminosity of young disk-star systems.} Numerical models predict that accretion dominates the total luminosity from the young disk-star system for a significant fraction of their early evolution, while measurements show that in the majority of cases, the observed accretion luminosity is less than 50\%\ of the bolometric one \citep[][]{2023ApJ...944..135F}. As discussed by these authors, one of the most plausible reasons for this discrepancy is that the protostars for which the accretion luminosity could be reliably measured typically have a low extinction and are therefore representative of a late evolutionary stage.

The most successful methodology to accurately separate stellar and accretion luminosity contributions in young disk systems is based on the assumption that the observational accretion signatures (e.g., hydrogen emission lines) are related to the total accretion luminosity as in Class~II objects so that the well-established Class~II correlations between the emission line and accretion luminosities \citep[e.g.,][]{2017A&A...600A..20A} can be applied to the earlier stages of evolution. 
{Correlations between line luminosities and accretion luminosities have been established empirically, so lines can be used as a quantitative proxy of accretion, even if the line formation region is not necessarily confined to the accretion flow. For example, the recent optical interferometry study of \citet{2024A&A...684A..43G} has found that the accreting material contributes only a small fraction of the total Br$\gamma$ line flux in Herbig stars.}
In the case of young embedded disks, the detection of near-infrared lines (especially \Brg) is possible \citep[e.g.,][]{2021A&A...650A..43F, 2011A&A...534A..32A}. The major limitation of   these studies is the very uncertain extinction correction, which is also the reason for choosing low-extinction Class~I objects as targets. 
More direct methods of measuring \Lacc\ using the optical excess emission have been developed \citep{2004ApJ...616..998W}, but again they are only applicable to a subset of objects with very low extinction and likely at the transition with the Class~II stage.

Younger and more embedded protostars require a method for deriving an accurate measurement of the extinction, which can be computed from the observed line ratios if the intrinsic values are known.
\citet{2007AJ....133.1673B} attempted to use \Bra\ and \Brg\ spectroscopy to derive extinction and accretion luminosity measurements for Class~I objects in Taurus. The methodology for deriving extinction in that study relies on the assumption that the intrinsic line ratio follows the Case~B theoretical ratio \citep[e.g.,][]{1987MNRAS.224..801H}. In this paper, we propose a different approach based on the assumption that the intrinsic line ratio for Class~I objects is the same as for Class~II objects, which we derive empirically from observations. Thus, \Lacc\ can then be computed from the extinction-corrected line luminosity using the correlations established for Class~II objects, as in \citet{2021A&A...650A..43F}.

There is an obvious advantage to using hydrogen lines at longer wavelengths where the extinction is much smaller.
The 4.05$\mu$m \Bra\ line is at least as strong as  \Brg\ \citep{1991ApJ...367..173G} and much less affected by uncertainty in the extinction estimates. As an example, for a young disk with \AV$= 50$~mag, the extinction at the \Brg\ wavelength is about 5~mag, while at the \Bra\ wavelength, it is only $\sim$2-3 mag \citep[depending on the adopted extinction law][see also Sect.~\ref{sec:extinction} and~\ref{sec:caveats_extinction}]{2019ApJ...877..116W}. However, this procedure requires characterization of the Class~II hydrogen line spectrum in the mid-infrared to establish the values of the line ratios and the correlations of \Lacc-\Lline, which are currently unavailable, with very limited exceptions \citep{2013ApJ...769...21S,2015ApJ...801...31R,2024A&A...684L...8R,2024AJ....167..232T}. The purpose of this paper is to present a pilot study of what
can be obtained from three hydrogen lines in the 2-4~$\mu$m region (\Brg (H(7-4), $\lambda=2.17 \mu$m), \Pfg (H(8-5), $\lambda=3.74 \mu$m), \Bra (H5-4), $\lambda=4.05 \mu$m). In this work, we first derive the mean values and their errors for two ratios (\Pfg/\Brg\ and \Bra/\Brg) and then establish the relations between \Lacc\ and \LPfg, and \Lacc\ and \LBra. 
Subsequently, we test how the Class II data can be used to derive the accretion luminosity in a sample of Class I objects and discuss the results.

We note that, more generally, this method can be used to derive the extinction correction for Class~II objects located in very dense and obscured regions of their parental molecular cloud. For these objects more direct approaches (e.g., directly measuring the excess continuum luminosity at UV and optical wavelengths) cannot be applied, and extinction plays a dominant role in computing the correct line luminosities.

\section {Observations and data reduction}
\label{sec:Obs}

The new spectra presented in this paper include eight Class II objects, seven T~Tauri stars (TTSs) and one Herbig Ae star (HAE), and 17 Class I and flat spectrum objects located in the Taurus and \roph\ star-forming regions. The objects are listed in Table \ref{tab:tab_classii} and \ref{tab:tab_classi_prop}, respectively.

The observations were performed using the SpeX instrument \citep[][with the post 2014 upgrade]{2003PASP..115..362R} mounted on the NASA IRTF telescope.
We used the {\tt LXD\_short} spectroscopic mode covering the wavelength range $1.67$-$4.2~\mu$m and the 0.5~arcsec wide and 15~arcsec long slit, resulting in a spectral resolution of R$\sim 1500$.
Observations were carried out in four separate sessions on 19-20 July 2022, 26-27 April 2023, 23-24 November 2023, and 27 December 2023.

We adopted the standard SpeX observing strategy: nodding on slit for all point-like calibrators and targets; executing instrument calibration sessions every $\sim 1.5$~hours; integrating on target sources for 15-30~minutes, depending on the target brightness. Finally, we observed A0V bright stars at different air masses every night for telluric correction.

We followed the standard SpeX data reduction and spectral extraction procedures using the IDL software package {\tt spextool} \citep{2004PASP..116..362C}.
After calibration and combination of all the exposures for each target and telluric standard, we extracted and combined the spectra and used {\tt spextool} to correct for telluric absorption. In order to check the final flux calibration, following the standard calibration steps, we imported the two spectral orders containing the \Brg\ line and the \Bra\ and \Pfg\ lines in a custom Python procedure, and we compared our final spectra with 2MASS and Spitzer photometry. Typically, we found the discrepancy to be less than $\sim$30\%, and we decided to scale the spectra to match the photometry. Using this approach, we minimized the uncertainty on the relative calibration of the two bands. A possible continuum variability would result in a corresponding uncertainty in the estimated value of the accretion luminosity.

Line fluxes were computed fitting a Gaussian line profile to the spectra.
The line fluxes measured for Class~II objects are reported in Table~\ref{tab:tab_classii}, while those measured in Class~I and flat spectrum sources are reported in Table~\ref{tab:tab_classi}. {We do not explicitly include the possible uncertainty due to the flux calibration procedure discussed above.}

\begin{table*}[]
    \caption{Class II SpeX data.}
    \label{tab:tab_classii}
    \centering
    \begin{tabular}{lcccccc}
    \hline
Name & d\tablefootmark{a} & A$_{\rm K}$ & F$_{Br\gamma}$ & F$_{Pf\gamma}$ &F$_{Br\alpha}$ \\
     &  (pc) & (mag) & ($10^{-17}\rm W/m^2$) & ($10^{-17}\rm W/m^2$)& ($10^{-17}\rm W/m^2$)\\
    \hline
BP Tau & 127 & 0.03\tablefootmark{b} & 13.2$\pm$0.5  & 3.1 $\pm$ 0.2& 14.6$\pm$0.3\\
CI Tau & 160 & 0.34\tablefootmark{b} & 24.2$\pm$0.4 & 10.1 $\pm$ 0.3& 24.6$\pm$0.4 \\
CW Tau & 131 & 0.47\tablefootmark{b} & 22.2$\pm$0.6 &11.2 $\pm$ 0.7& 32.0$\pm$0.8\\
SSTc2d J162816.7-240514 & 138 & 0.78\tablefootmark{c} & 1.0 $\pm$0.2 & <1& <2\\
SSTc2d J162755.6-242618 & 132 & 0.13\tablefootmark{c} &   4.6$\pm$0.2& 1.2 $\pm$ 0.2 & 5.5$\pm$0.4 \\
SSTc2d J162556.1-242048 & 135 & 0.35\tablefootmark{c} &   34.4$\pm$0.7& 10.8 $\pm$ 0.7&26.9$\pm$1.0\\
SSTc2d J162617.1-242021 & 135 & 0.39\tablefootmark{c} &   7.3$\pm$0.4& <1.8& 7.1$\pm$0.7\\
HD163296 & 101 & 0.0\tablefootmark{d} & 215$\pm$6.8 &87.5 $\pm$ 6.8&  278.1$\pm$8.6\\
    \hline
    \end{tabular}
\tablefoot{  \\ 
\tablefoottext{a}{Distances are from Gaia EDR3 \citep{2021A&A...649A...1G}.}
\tablefoottext{b}{Extinction from \citet{Gangi2022} computed from A$_{\rm V}$ using \citet{Cardelli1989} with R$=3.1$.}
\tablefoottext{c}{Extinction from \citet{Testi2022}.}
\tablefoottext{d}{Extinction from \citet{UbeiraGabellini2019}.}
}
\end{table*}

\begin{table*}[]
    \caption{Properties of the newly observed Class I/F and those from  \citet{2007AJ....133.1673B} (below the horizontal line).}
    \label{tab:tab_classi_prop}
    \centering
    \begin{tabular}{llllllllll}
    \hline
&Name & $\alpha_{2000}$ & $\delta_{2000}$& Region & d\tablefootmark{a} &  Class\tablefootmark{b} & A$_{\rm K,fg}$\tablefootmark{c}  &T$_{\rm bol}$\tablefootmark{c} & L$_{\rm bol}$\tablefootmark{c} \\
&    &          &        &       & (pc) &  & (mag)  & (K) & (L$_\odot$) \\
    \hline
1&SSTc2d j162621.3-242304 &16 26 21.38 &-24 23 04.0 & Oph L1688&139.4 &  I & 2.2 & $320^{+137}_{-64}$ & $18^{+7}_{-4}$ \\
2&SSTc2d j162702.3-243727& 16 27 02.33& -24 37 27.2& Oph L1688&139.4  & I& 3.1 & $939^{+511}_{-262}$ & $16^{+18}_{-6}$\\
3&SSTc2d j162709.4-243718&16 27 09.43 &-24 37 18.8& Oph L1688&139.4 &  I& 1.7 & $455^{+80}_{-53}$ & $29^{+14}_{-9}$ \\
4&SSTc2d j162721.7-242953& 16 27 21.80 & -24 29 53.4& Oph L1688&139.4 & I& 2.0 & $534^{+67}_{-54}$& $2^{+1}_{-0.6}$ \\
5&SSTc2d j162726.9-244050 & 	16 27 26.91& -24 40 50.8 & Oph L1688& 139.4 & I& 2.5 & $408^{+117}_{-81}$ & 6$^{+2}_{-1.5}$ \\
6&SSTc2d j162727.9-243933 &16 27 27.89 & -24 39 43.9& Oph L1688&139.4 & I& 2.2 & $314^{+45}_{-26}$& $13^{+5}_{-3.5}$ \\
7&SSTc2d j162730.1-242743 & 16 27 30.18& -24 27 43.4& Oph L1688&139.4 & F& 2.2 & $814^{+288}_{-171}$ & $3^{+2.5}_{-1}$ \\
8&SSTc2d j163135.6-240129 & 16 31 35.66 & -24 01 29.5& Oph L1709 &139.4 &  F& 1.1 & $387^{+168}_{-110}$ & $2.5^{+0.9}_{-0.5}$ \\
9&SSTc2d j163200.9-245642 &16 32 00.99 &  	-24 56 42.0& Oph L1689 &139.4 & I& 1.1 & $133^{+45}_{-27}$ & $6^{+0.7}_{-0.5}$ \\
10&IRAS 04016+2610  & 4 04 43.07 & +26 18 56.4& Tau F&  146   & I & 0.4 & $427^{+81}_{-22}$ & $19^{+8.5}_{-2.4}$ \\
11&MHO 1 &4 14 26.27 & +28 06 03.3& Tau A  &130.5 & I & 0.9 & $1064^{+6330}_{-562}$ & $23^{+3900}_{-8}$ \\
12&MHO 2 &4 14 26.40 &+28 05 59.6& Tau A & 130.5 & I & 0.9 & $1136^{+6206}_{-604}$ & $23^{+4000}_{-8}$ \\
13&SSTtau 041430.5+280514 & 4 14 30.55& +28 05 14.6& Tau A  &130.5 & F  & 1.0 & $2490^{+5000}_{-1860}$ & $25^{+1100}_{-15}$ \\
14&SSTtau 042107.9+270220 & 4 21 07.97 & +27 02 20.1& Tau A & 130.5  & I  & 2.7 & $1284^{+921}_{-556} $ & $17^{+38}_{-8}$ \\
15&SSTtau 042111.4+270109 & 4 21 11.47 &+27 01 09.4& Tau A & 130.5 & I  & 2.7 & $1274^{+1015}_{-443}$ & $8^{+19}_{-4}$\\
15&SSTtau 043232.0+225726 & 4 32 32.05 &+22 57 26.7 & Tau B & 131 &  I  & 2.1 & $838^{+719}_{-349}$ & $9^{+11}_{-3}$\\
17& SSTtau 043935.1+254144 &4 39 35.19 &+25 41 44.7& Tau B & 131 &  I  & 0.4 & $337^{+54}_{-19}$ & $13^{+4.6}_{-1.4}$ \\
\hline
18& IRAS04108+2803B &4 13 54.72&+28 11 32.9 & Tau A&130.5 & I & 0.9 & $270^{+110}_{-55}$ & $6.0^{+1.8}_{-1}$ \\
18& IRAS04158+2805 &4 18 58.13& +28 12 23.4&Tau A &130.5 & I & 0.6 & $1172^{+6058}_{-700}$ & $2^{+3800}_{-0.6}$ \\
20& IRAS04181+2654A & 4 21 11.47& +27 01 09.4& Tau E& 160.2 & F & 1.1 & $434^{+235}_{-124}$ & $8^{+3.5}_{-1.9}$ \\
21&IRAS04239+2436 &4 26 56.30& +24 43 35.3&Tau B &131 & I & 0.4 & $311^{+961}_{-28}$ & $13^{+7}_{-1.3}$ \\
22&IRAS04295+2251 &4 32 32.05&+22 57 26.7 &  Tau E &160.2 & F & 2.1 & $1544^{+3060}_{-1016}$ & $23^{+6770}_{-11}$ \\
23&IRAS04361+2547 & 4 39 13.89& +25 53 20.9& Tau B  &131 & I & 0.4 & $163^{+36}_{-10}$ & $28^{+5.5}_{-1.8}$ \\
24&IRAS04365+2535 &4 39 35.19 & +25 41 44.7& Tau B &131 & I & 0.4 & $369^{+89}_{-27}$ & $11^{+4.3}_{-1.2}$\\

\hline
    \end{tabular}
\tablefoot{  \\ 
\tablefoottext{a}{ Distances are the average cloud distance for Ophiuchus and average distances of each  sub-cloud for Taurus \citep[see][]{Testi2022,2021A&A...652A...2G,2020A&A...638A..85R}.}
\tablefoottext{b}{ The literature young stellar object classification is based on \citet{2015ApJS..220...11D}, \citet{2010ApJS..186..259R}, and \citet{2007AJ....133.1673B}.}
\tablefoottext{c}{From this paper (see Sec.~\ref{sec:lbol_tbol}).}
}
\end{table*}
\begin{table*}[]
    \caption{Class I accretion properties.}
    \label{tab:tab_classi}
    \centering
    \begin{tabular}{llcccccccc}
    \hline
&Name &  F$_{\rm Br\gamma}\tablefootmark{a}$      & F$_{\rm Pf\gamma}\tablefootmark{a}$        & F$_{\rm Br\alpha}$ \tablefootmark{a}      &  A$_{\rm Pf\gamma}$\tablefootmark{b}&  A$_{\rm Br\alpha}$\tablefootmark{c}& Log$_{\rm 10}$L$_{\rm acc}$\tablefootmark{d}\\
&     & ($10^{-17}\rm W/m^2$)& ($10^{-17}\rm W/m^2$) &($10^{-17}\rm W/m^2$) &(mag)   &(mag)        &  (L$_\odot$)&\\

    \hline
1&SSTc2d j162621.3-242304 & 5.3$\pm$ 0.5    & <7             & 26.3$\pm$4.0& <2.1 & 2.1$\pm$ 0.6  &0.80$\pm$ 0.59\\
2&SSTc2d j162702.3-243727 & 18.5 $\pm$ 0.5  & 37.6 $\pm$ 1.8 &133.9$\pm$ 2.8& 2.8$\pm$0.4&  2.6$\pm$0.4 & 1.89 $\pm$0.39\\
3&SSTc2d j162709.4-243718 & 20.1 $\pm$ 0.6  &  68$\pm$ 11 &209$\pm$ 30& 3.6$\pm$0.5& 3.0$\pm$ 0.5& 2.35 $\pm$0.51\\
4&SSTc2d j162721.7-242953 & <1.0            & <2             &11.1$\pm$  1.1& --& --&-- \\
5&SSTc2d j162726.9-244050 & 1.3 $\pm$ 0.2   & <3             &15.0$\pm$  0.9& <3.0 & 3.2$\pm$ 0.5& 1.01$\pm$ 0.51\\
6&SSTc2d j162727.9-243933 &2.0$\pm$ 0.2     &<5              &16.54$\pm$ 1.7& <3.1& 2.8$\pm$ 0.5& 0.86$\pm$ 0.54\\
7&SSTc2d j162730.1-242743 & 0.6$\pm$ 0.16   & <3             & <8.2       &-- &--  &--\\
8&SSTc2d j163135.6-240129 &1.0 $\pm$ 0.1&  2.4 $\pm$0.4     &12.9$\pm$ 0.8 &3.1$\pm$0.6 &3.32$\pm$0.5& 0.99$\pm$ 0.50\\
9&SSTc2d j163200.9-245642  &<0.7 & <3                         &9.8 $\pm$  1.5&-- &--&-- \\
10& IRAS 04016+2610 &<3 & <4                           & 9.8$\pm$  1.6&--& --& --\\
11&MHO 1& 11.1 $\pm$ 0.6&6.1$\pm$0.6&24.0$\pm$  1.0 &0.9$\pm$0.6 &1.1$\pm$0.4& 0.16$\pm$ 0.48\\
12&MHO 2 &9.0$\pm$ 0.9&  <3         &8.63$\pm$ 0.9& <0.2 & 0.1$\pm$ 0.5& -0.88 $\pm$0.59\\
13&SSTtau 041430.5+280514 & 3.9 $\pm$ 0.6 & <2        &6.8 $\pm$ 0.5& <0.5& 0.8$\pm$ 0.5&. -0.65 $\pm$0.58\\
14& SSTtau 042107.9+270220 &0.9 $\pm$ 0.06& 0.7 $\pm$ 0.1     & 5.1$\pm$0.06& 1.4$\pm$0.6 &2.2$\pm$ 0.4 &-0.11$\pm$ 0.47\\
15&SSTtau 042111.4+270109 & 1.8 $\pm$0.15& 1.0 $\pm$0.1&2.8$\pm$  0.2& 0.9$\pm$0.5& 0.67$\pm$0.5& -1.20 $\pm$0.56\\
16& SSTtau 043232.0+225726& 1.1$\pm$0.13 &0.9$\pm$0.2& 2.4$\pm$  0.3&  1.4$\pm$0.7& 1.1$\pm$ 0.5& -1.05 $\pm$0.62\\
17&SSTtau 043935.1+254144 & 1.4 $\pm$0.07 &3.4$\pm$0.3&13.5$\pm$ 0.5& 3.1$\pm$0.5&  3.0$\pm$ 0.4& 0.79 $\pm$0.46\\
\hline
18&IRAS04108+2803B & 0.3$\pm$0.1&  2.8$\pm$0.1 &9.6$\pm$ 0.4&4.9$\pm$0.5& 4.3$\pm$ 0.5& 1.25$\pm$ 0.48\\
19&IRAS04158+2805 &  0.2$\pm$0.06&     <0.4    & 1.0$\pm$  0.2& <3 & 2.3$\pm$ 0.8& -0.96 $\pm$0.79\\
20&IRAS04181+2654A& 0.8$\pm$0.1  &1.5$\pm$0.3  & 5.3$\pm$  0.3&2.6$\pm$0.7& 2.44$\pm$0.5& 0.23$\pm$ 0.54\\
21&IRAS04239+2436&  5.3$\pm$0.06 &11.7$\pm$0.2 &31.9$\pm$ 0.4& 2.9$\pm$0.3& 2.4$\pm$ 0.4& 0.94 $\pm$0.40\\
22&IRAS04295+2251& 0.4$\pm$0.1   &<0.3         & 1.8$\pm$  0.2& <1.4& 2.0$\pm$ 0.7& -0.55$\pm$  0.66\\
23&IRAS04361+2547& 0.9$\pm$0.06 &  1.0$\pm$0.1 &2.8$\pm$  0.08& 1.9$\pm$0.5& 1.5$\pm$ 0.4& -0.78$\pm$ 0.50\\
24&IRAS04365+2535& 1.3$\pm$0.1 & 2.7$\pm$0.3   &13.1$\pm$ 0.2& 2.9$\pm$0.5&3.02$\pm$0.4& 0.79$\pm$ 0.45\\
\hline
 \end{tabular}
\tablefoot{  \\ 
\tablefoottext{a}{ Observed fluxes: Objects 1--17 are from this paper, objects 18--24 are from Beck.}
\tablefoottext{b}{ Extinction at the \Pfg\ wavelength from the ratio \Brg/\Pfg.}
\tablefoottext{c}{ Extinction at the \Bra\ wavelength from the ratio \Bra/\Brg.}
\tablefoottext{d}{Accretion luminosity computed from the \Bra\ luminosity (eq.(2)).}
}
\end{table*}

\section {Class II}
\label{sec:results_classII}

As outlined in Sec.~\ref{sec:Intro}, we intended to use the line measurements on the Class II objects for two purposes. The first goal was to measure the fluxes of the ratios \Pfg/\Brg\ and \Bra/\Brg and derive their mean values, which  we used to infer the extinction in highly reddened objects. The second was to extend the  correlation between optical and near-infrared  lines and the accretion luminosity to lines at longer wavelengths.

\subsection {Class II sample}
\label{discussion_classII_sample}
To increase the number of stars and extend our sample toward objects of higher \Lacc, we included the objects from
\citet{Nisini95}, who observed the same infrared lines in a sample of intermediate-mass pre-main sequence stars. Our final Class II sample includes a few TTSs with a high \Lacc\ and a number of HAeBe objects. This choice was motivated by the need to cover a line luminosity range comparable to what is expected for very young objects, such as Class I sources, which are known to have a higher \Lacc\ than the older TTSs in well-studied nearby star-forming regions \citep{2016ARA&A..54..135H}.  We note, however,
that this implies that the mass of the central stars in our Class~I sample may differ from those in the Class~II sample (the latter typically being of higher mass). 
This is probably not a problem, as the correlation between \Lacc\ and the luminosity of individual hydrogen lines (L$_{lines}$) does not significantly depend on the mass of the central star, with the possible exception of the very low mass stars and brown dwarfs (see Appendix~\ref{sec:app_lacc_lbrg}).

\subsection{Class II line ratios}
\label{Sec:ClassII_ratios}

We proceeded by first deriving the intrinsic ratio  of the \Bra/\Brg and \Pfg/\Brg\ fluxes for the Class II objects observed with SpeX. We corrected the observed fluxes for extinction using the values tabulated in Table~\ref{tab:tab_classii} and the \citet{Cardelli1989} extinction law with \RV=3.1.
For all stars in our sample, the observations of the three lines were either simultaneous (in the case of our SpeX observations) or executed within a few hours, such as in the case of the observations reported by \citet{Nisini95}. The total number of Class II objects is 25, and they all have detected \Brg\ lines. One has an upper limit in \Bra, and two have upper limits in \Pfg. 

\begin{figure*}
    \centering
    \includegraphics[width=5.9cm]{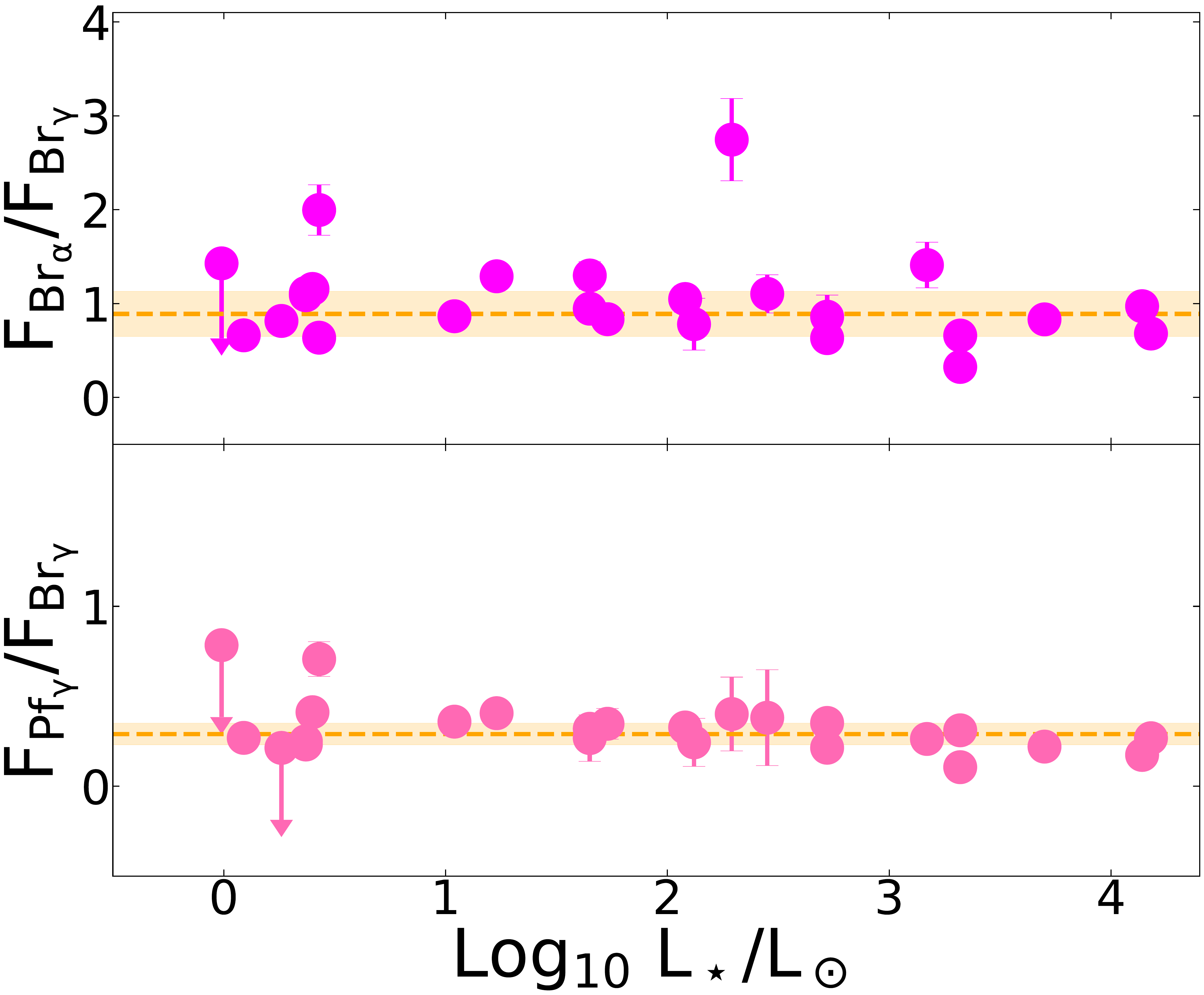}
    \includegraphics[width=5.5cm]{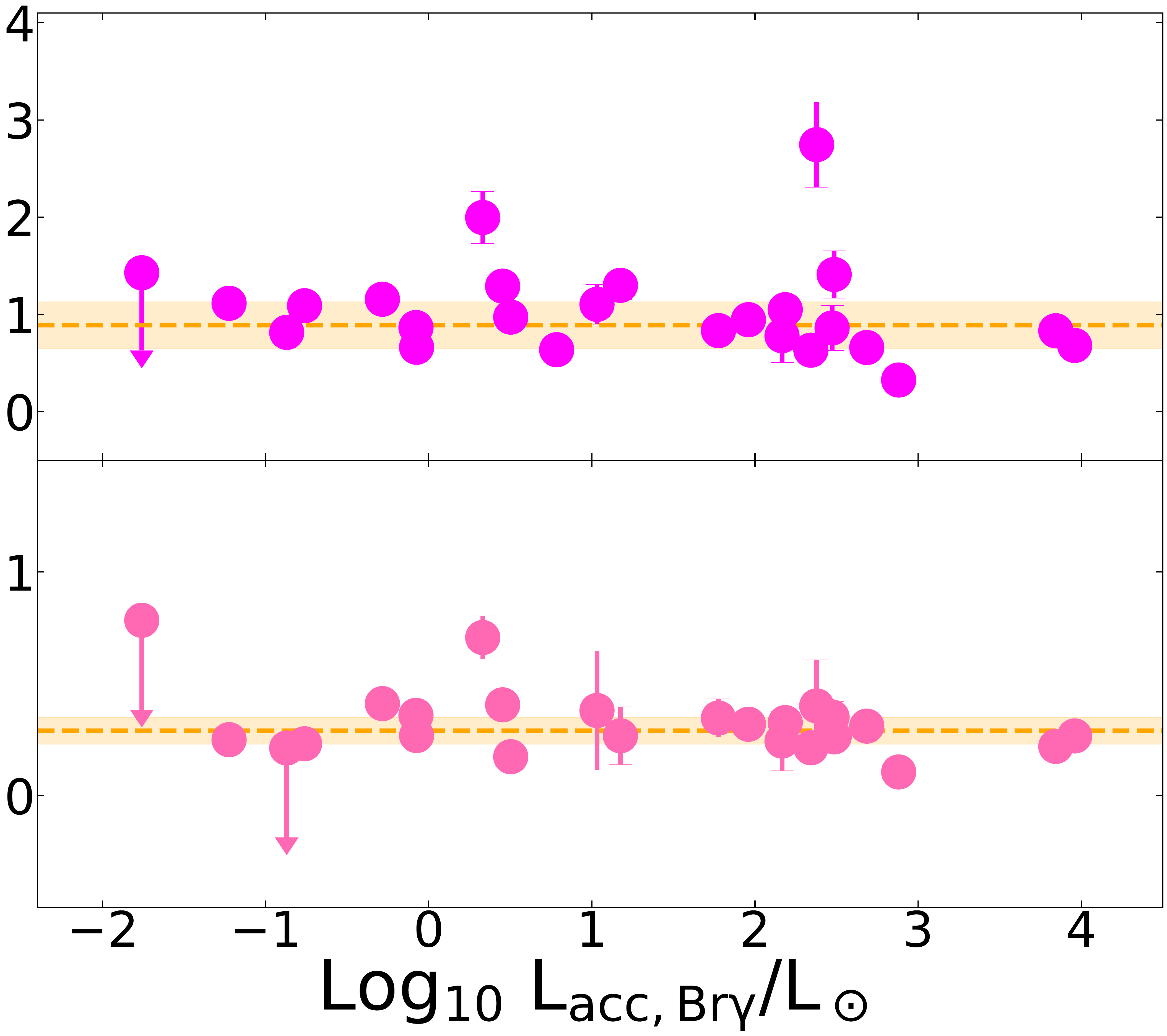}
     \includegraphics[width=5.5cm]{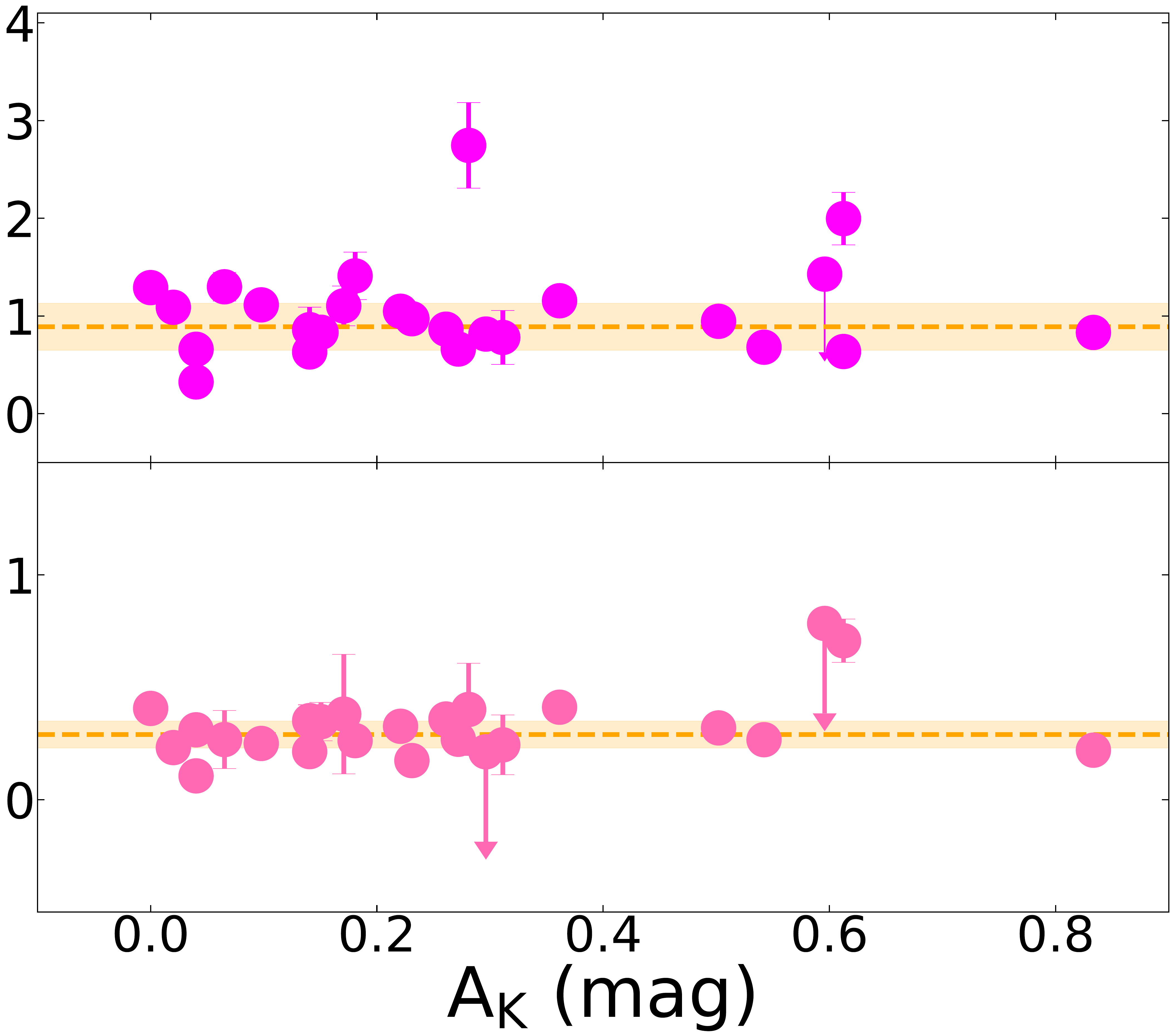}
    \caption{Top panel: Ratio of the de-reddened Br$\alpha$ and Br$\gamma$  fluxes of Class~II sources shown as a function of \Lstar\ (left), \Lacc, computed from the \Brg\ luminosity (center; see Sec.~\ref{sec:ClassII_lline}), and  the extinction in the K band (right). Bottom panel: Same but for the ratio of \Pfg\ to \Brg\ fluxes.  The values of the ratios are shown by filled dots; the 3$\sigma$ upper limits are indicated by arrows. When not visible, the errors are smaller than the dots. In both sets of panels, the mean values of the ratios and their uncertainties are shown in orange.}
    \label{fig:f_ratio_ClassII}
\end{figure*}

\begin{figure}
   \centering
   \includegraphics[width=9.0cm]{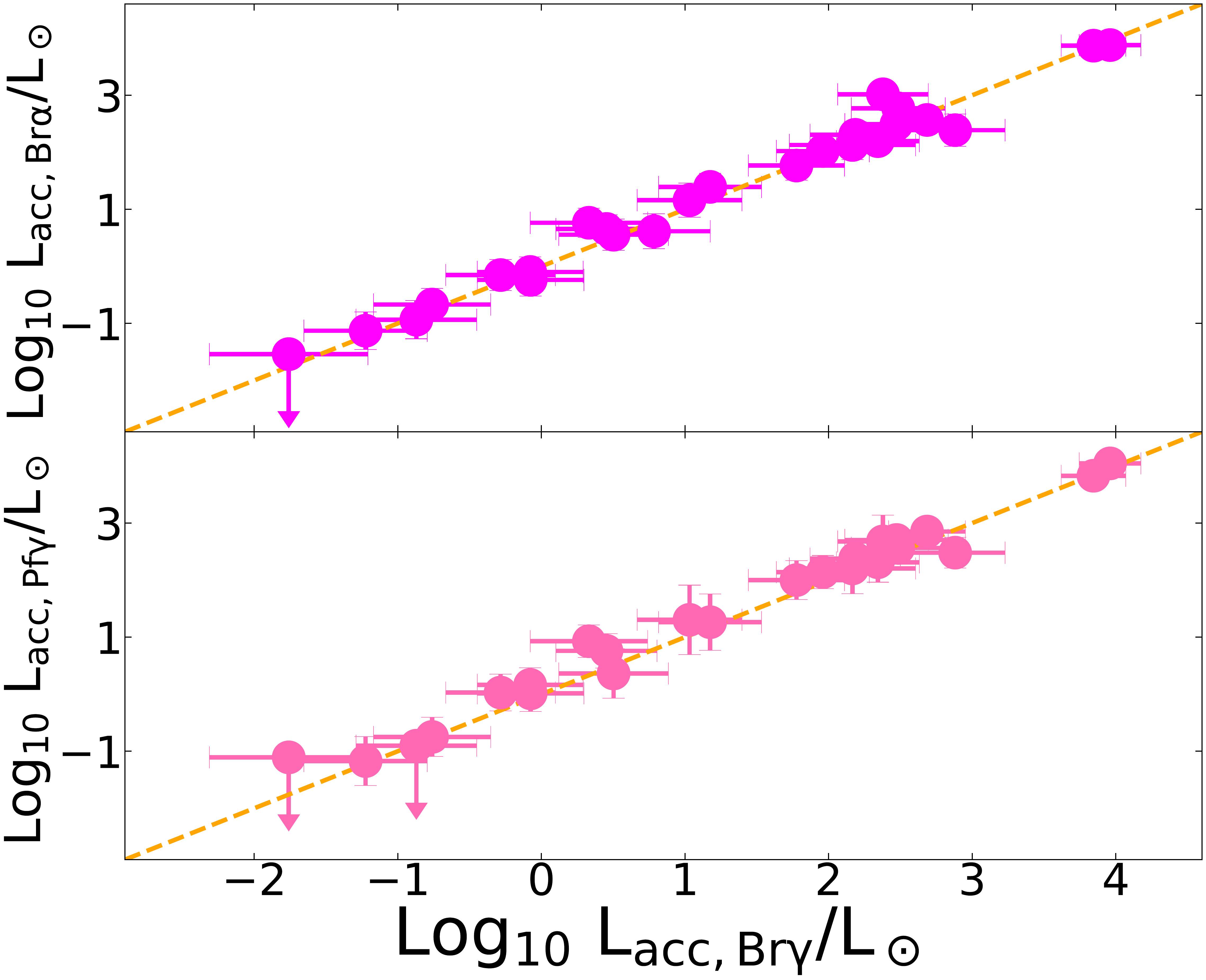}

   \caption{Top panel: Values of the accretion luminosity, \Lacc, derived from \LBra\ as a function of \Lacc\ derived  from \LBrg. Bottom panel: Same but for \Lacc\ derived from \LPfg. The line of equal values is shown in orange in each panel.}
    \label{fig:f_Lacc_comparison_ClassII}
\end{figure}

The ratios of the dereddened fluxes are shown in Fig.~\ref{fig:f_ratio_ClassII}.  
Each of the ratios cluster around a well-defined constant value independent of source properties, such as the stellar luminosity \Lstar, accretion luminosity, and extinction (see Fig.~\ref{fig:f_Lacc_comparison_ClassII}). The mean value and its error  
are respectively \rbra=0.89$\pm$0.24 and \rpfg=0.29$\pm$0.06.  These values are briefly discussed in Sec.~\ref{sec:Discussion_ClassII} in the general context of hydrogen line emission in Class~II objects.

\begin{figure}
    \includegraphics[height=7cm]{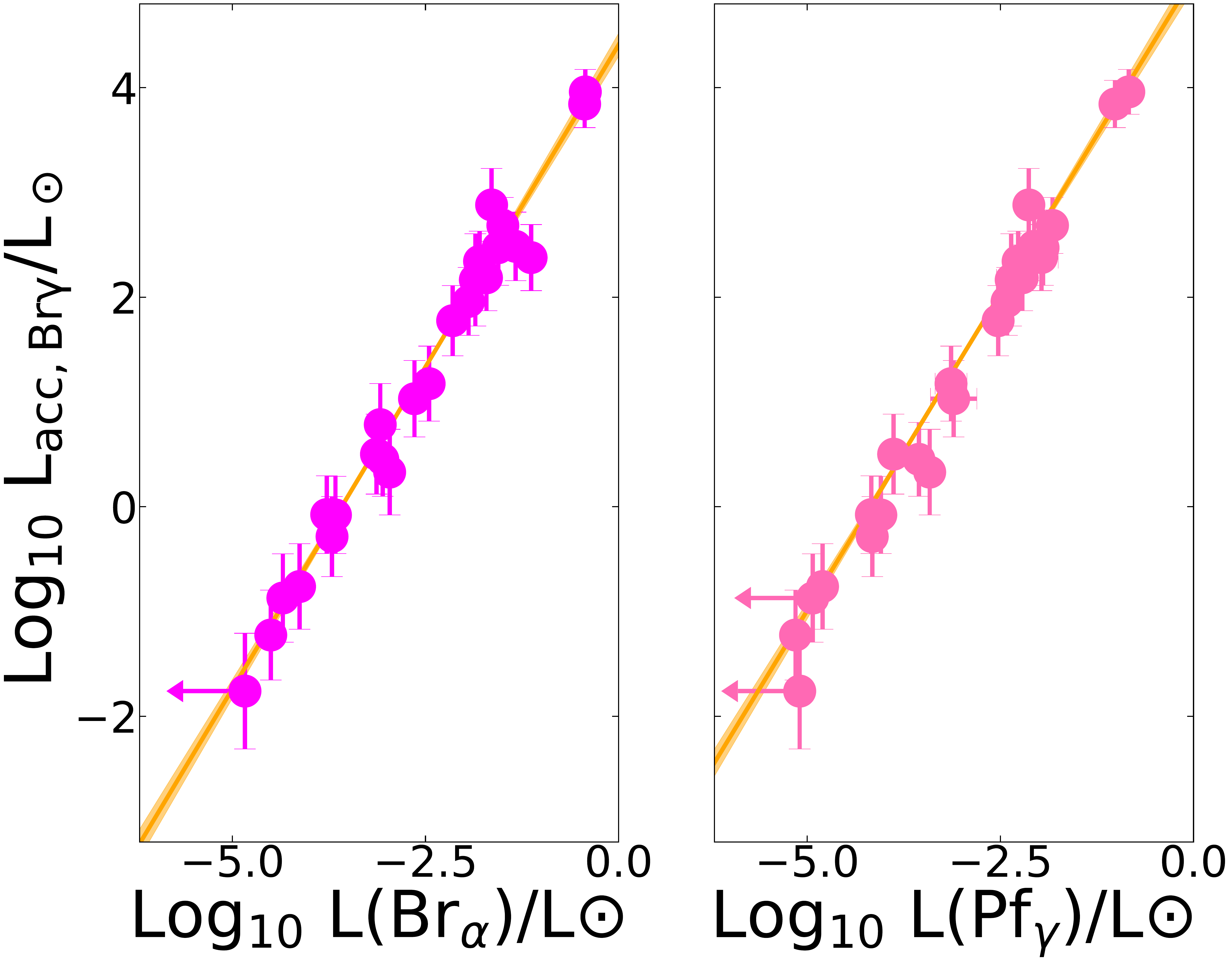} 
   \caption{Accretion luminosity, \Lacc, derived from the \Brg\ luminosity (eq.(1)) plotted as a function of the line luminosity for \Bra\ (left panel) and \Pfg\ (right panel), respectively. The orange line and shadowed area show the best-fitting correlation and their uncertainty (see eq.(2) and (3)). }
    \label{fig:f_lacc_lines_classii}
\end{figure}

\subsection{Class II \Lacc-\Lline\ relations}
\label{sec:ClassII_lline}

We derived the accretion luminosity, \Lacc, of each object using the well-established existence of a correlation between \Lacc\ and \LBrg\  \citep[e.g.,][]{1998AJ....116.2965M,2006A&A...452..245N,2017A&A...600A..20A,2017MNRAS.464.4721F}, revised in 
Appendix~\ref{sec:app_lacc_lbrg} to include only objects  with a high \LBrg\  and \Lacc\ derived from  UV excess emission:

\begin{equation}
\label{eq:lacc-lbrg}
{\rm Log}_{10}({\rm L}_{\rm acc}/{\rm L}_\odot) = (1.21\pm 0.05)\, {\rm Log}_{10}({\rm L}_{Br_\gamma}/{\rm L}_\odot) \, + (4.28\pm 0.18).
\end{equation}

No similar datasets exist for \Bra\ and \Pfg\ in the literature. For our sample, we do not directly have the values of \Lacc\ computed from a measured UV excess. Therefore, we computed the correlations of these two lines with \Lacc as derived from \LBrg. In practice, we relied on the sample of Class II objects in Table ~\ref{tab:tab_classi}. We computed   \Lacc\  from  \LBrg\ and eq.(1), and we derived the correlation between \Lacc\ and \LBra\ (and \LPfg). The uncertainties in the determination of the linear fit coefficients of Eq.~\ref{eq:lacc-lbra} and~\ref{eq:lacc-lpfg} were obtained by propagating the errors on the measured line fluxes and on the \Lacc-\LBrg\ relation used to determine \Lacc. The results are (see Fig.~\ref{fig:f_lacc_lines_classii})

\begin {equation}
\label{eq:lacc-lbra}
{ \rm Log_{10}  (L_{\rm acc}/L_\odot)=(1.23\pm 0.04)} \times {\rm Log_{10} (L_{Br_\alpha}/L_\odot)} + (4.41 \pm 0.11)
\end{equation}

and
\begin{equation}
\label{eq:lacc-lpfg}
{ \rm Log_{10}  (L_{acc}/L_\odot)=(1.21\pm 0.04)} \times {\rm Log_{10} (L_{Pf_\gamma}/L_\odot)} + (5.06 \pm 0.12).
\end{equation}
Figure~\ref{fig:f_Lacc_comparison_ClassII} shows the values of \Lacc\ derived from \LBra\ (top panel) and \LPfg\ (bottom panel) plotted as a function of the accretion luminosity derived from \LBrg. 
The agreement between the values from \LBra\ and \LPfg\ is very good and well within the uncertainties.

\section{Class I objects}
\label{sec:ClassI}

In this section, we apply the results obtained for our  Class II sample to derive the extinction toward each Class I object and then its accretion luminosity. 
\citet{2007AJ....133.1673B} observed a sample of Class~I sources in Taurus with the same instrument and setup we used. We include in our analysis the seven objects for which they detected  Br$\gamma$ and at least one among Pf$\gamma$ and Br$\alpha$. For homogeneity, we re-analyzed all of the objects using the same approach employed in our sample.

\subsection{Extinction}
\label{sec:extinction}

\begin{figure}
    \centering
    \includegraphics[width=9.0cm]{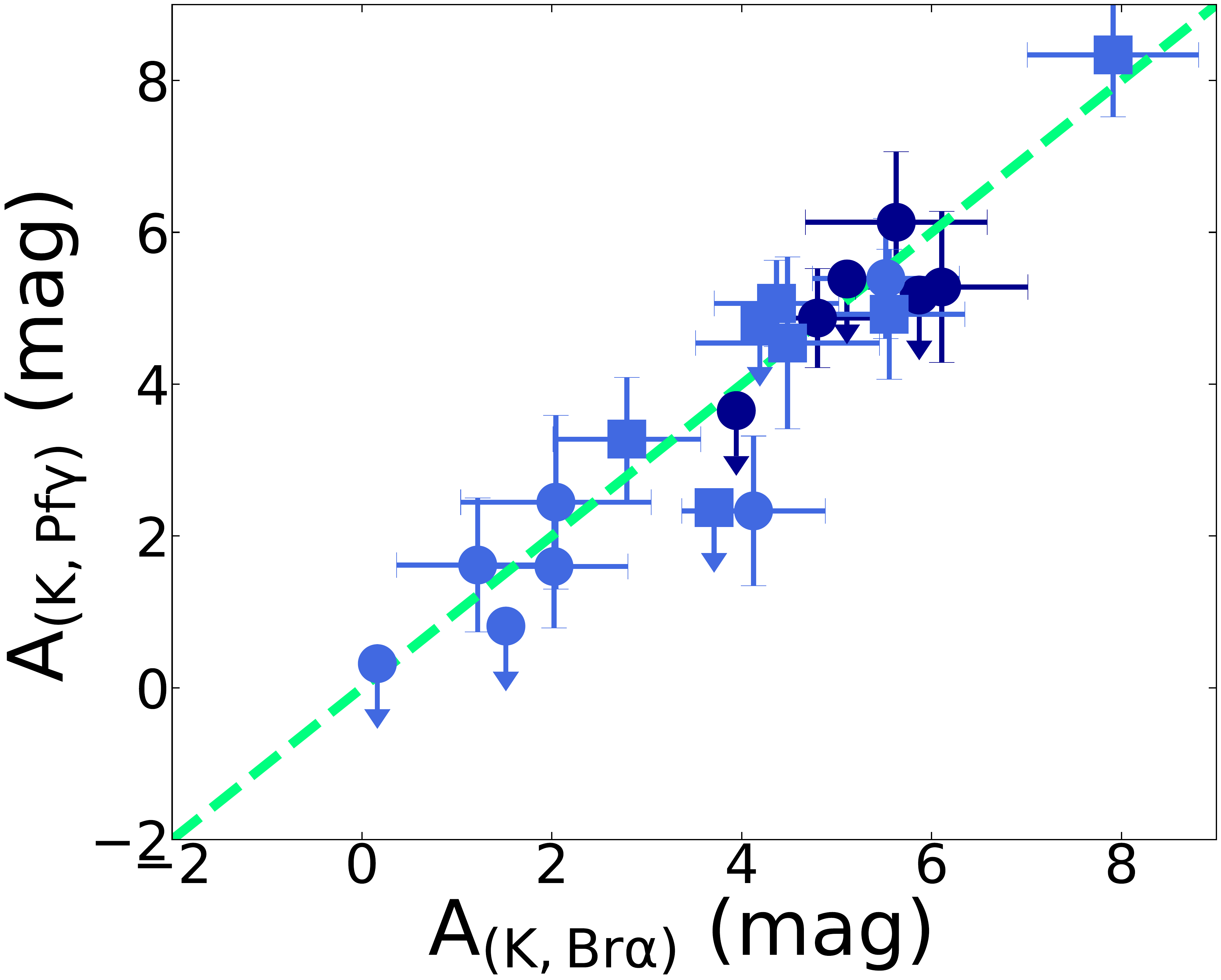} 
    \caption{Accretion luminosity computed from \Pfg\ shown as a function of the accretion luminosity computed from \Bra\ for Class~I objects. In the first case, the extinction is derived from the ratio \Pfg/\Brg, and in the second it is from \Bra/\Brg (see Sect.~\ref{sec:extinction}). Dark blue dots are objects in Ophiuchus; light blue dots are objects in Taurus. Light blue squares are Taurus objects from \citet{2007AJ....133.1673B}. The green dashed line shows the locus of equal values.}
    \label{fig:f_AKbra_AKpfg}
\end{figure}

The extinction in any of the three lines can be obtained from the ratio of two of them, once an extinction law is adopted.
We computed the optical depth toward the Class I objects at the \Bra\ wavelength  from the observed flux ratio R$_{\rm{obs}}$=\Bra/\Brg, assuming that the intrinsic  ratio is the same as in Class II objects:
\begin{equation}
  \tau (\rm Br_\alpha)=\>\> \ln \>\>\bigg( \frac {R_{obs}}{R_{II}} \bigg )\>\times \> \bigg(\frac{1}{{ A_{2.17}/A_{4.05}-1}}\bigg).
\end{equation}
The extinction is  A$_\lambda$= $1.086\, \tau_\lambda$. In Eq.(4),  $\rm {R_{II}}$ is the intrinsic \Bra/\Brg\ extinction-corrected ratio for Class~II objects ($\rm {R_{\rm II}}$=0.89$\pm$0.24, see Sect.~\ref{Sec:ClassII_ratios}), and ${\rm A_{2.17}/A_{4.05}}$ is the ratio of the extinction at the wavelength of the lines ($\rm A_\lambda=1.086\times\tau_\lambda$).
We adopted in all cases the extinction law WD01 with \RV=5.5 \citep{2001ApJ...548..296W}, as it is more suitable for the dense regions surrounding  Class I sources \citep[appropriate for \AK$\ge 2.0$, also according to][]{2023Univ....9..364L}; the corresponding ratio is ${\rm A_{4.05}/A_{2.17}} = 0.534$. We proceeded in the same way to derive the optical depth at the \Pfg\ line from  the observed \Pfg/\Brg\ flux ratio:
\begin{equation}
  \tau (\rm Pf_\gamma)=\>\> \ln \>\>\bigg( \frac {R_{obs}}{R_{II}} \bigg )\>\times \> \bigg(\frac{1}{{ A_{2.17}/A_{3.74}-1}}\bigg),
\end{equation}
where  R$_{\rm{obs}}$=\Pfg/\Brg, $\rm{R_{II}}$=0.29$\pm$0.06 for \Pfg/\Brg (Sect.~\ref{Sec:ClassII_ratios}), and ${\rm A_{3.74}/A_{2.17}} = 0.573$.

The extinctions derived for \Bra\ and \Pfg\ are given in Table~\ref{tab:tab_classi}. While the two extinctions were computed independently, we checked that the corresponding extinctions in K are equal within the uncertainty (see Fig.~\ref{fig:f_AKbra_AKpfg}).
The values of the extinction for the sources in our sample are in the range of \AK$\sim 1-8$ mag (\AV$\sim 10-80$ mag).

\subsection{Accretion luminosity}

The accretion luminosity of each object was computed from the luminosity of \Bra\ and \Pfg\, corrected for extinction and as derived in the previous section using  the correlation between the line luminosity and \Lacc\ derived in Sec.~\ref{sec:ClassII_lline}.  
\begin{figure}
    \centering
    \includegraphics[width=9.0cm]{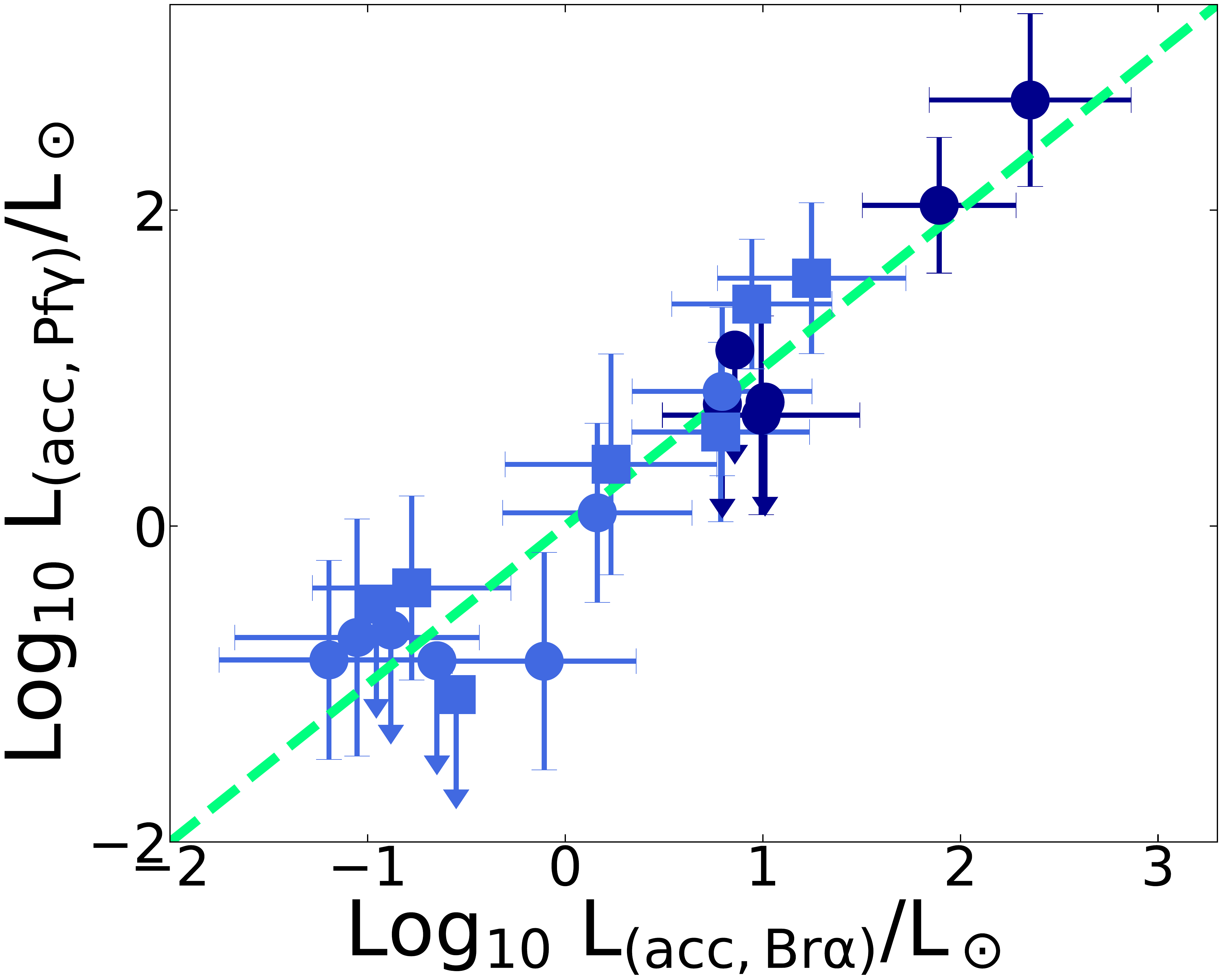} 
    \caption{Extinction in the K band computed from the ratio \Pfg/\Brg\ is shown versus the value derived from the  \Bra/\Brg\ ratio ( see Sect.~\ref{sec:extinction}). Symbols are as in Fig.~\ref{fig:f_AKbra_AKpfg}. The green dashed line shows the locus of equal values.}
    \label{fig:f_Lbra_Lpfg}
\end{figure}
The two values agree well within uncertainties, as shown in Fig.~\ref{fig:f_Lbra_Lpfg}. We note that the errors include the contributions from the observed line fluxes, the extinction at the wavelength of the line we used to measure \Lacc, and the uncertainty on the relation \Lacc-\LBra\ (or \Lacc-\LPfg). The values displayed in Table~\ref{tab:tab_classi} are derived from the \Bra\ line, as this line is detected in all objects and the values of \Lacc\ derived from it are more accurate than the value derived from the other two lines. These \Lacc\ values are the ones we use in the rest of the paper, and we refer to them as 
\Lacc\ for simplicity.

\subsection{Foreground extinction, bolometric luminosity, and temperature}
\label{sec:lbol_tbol}
  
To compute the bolometric luminosities (\Lbol) and bolometric temperatures (\Tbol) of all of our sources, we followed the general methodology of \citet{2015ApJS..220...11D} but with the important difference that we tried to evaluate the foreground extinction for each of our sources independently (instead of using full cloud averages). For each target, we selected an area with a radius of 150$^{\prime\prime}$, or about 0.1~pc at the distance of our targets, and we computed the average extinction of all Class~II within this area \citep[from the catalogs of][for Taurus and Ophiuchus, respectively]{2019AJ....158...54E,2020AJ....159..282E}.
The choice of the maximum separation was based on ensuring that a few Class~II stars are available to compute an average extinction for each Class~I object.
Two of our objects (SSTc2d~J163135.6$-$240129 and SSTc2d~J163200.9$-$245642) 
are outside the area surveyed by \citet{2020AJ....159..282E}, and for these we adopted the average cloud extinction value of \citet{2015ApJS..220...11D}. 

We decided to use the average foreground extinction using Class~II stars within $\sim 0.1$~pc around each of our targets, instead of the average cloud extinction, because we noticed that Class~II sources located in the vicinity of our targets show a significantly higher extinction than average. This is not surprising, as Class~I sources are typically found in the denser regions of molecular clouds, while the full Class~II population is much more dispersed \citep{2009ApJS..181..321E}, leading to lower average foreground extinction.
The average uncertainty of our computation of the foreground extinction is about $\Delta\rm A_{\rm K}\sim 1$~mag. 
The derived values of A$_{\rm K,fg}$, \Lbol, and \Tbol\ are listed in Table~\ref{tab:tab_classi_prop}. Compared to the estimates of \citet{2015ApJS..220...11D}, our values of \Lbol\ and \Tbol\ are generally higher, by up to a factor of about two, for objects with the largest values of A$_{\rm K,fg}$. In the range of A$_{\rm K,fg}$ of our sample ($\sim 0.5-3.2$), we found no correlation between the foreground extinction and the derived \Lbol\ and \Tbol, as expected. 

\begin{figure}
    \centering
    \includegraphics[width=9.0cm]{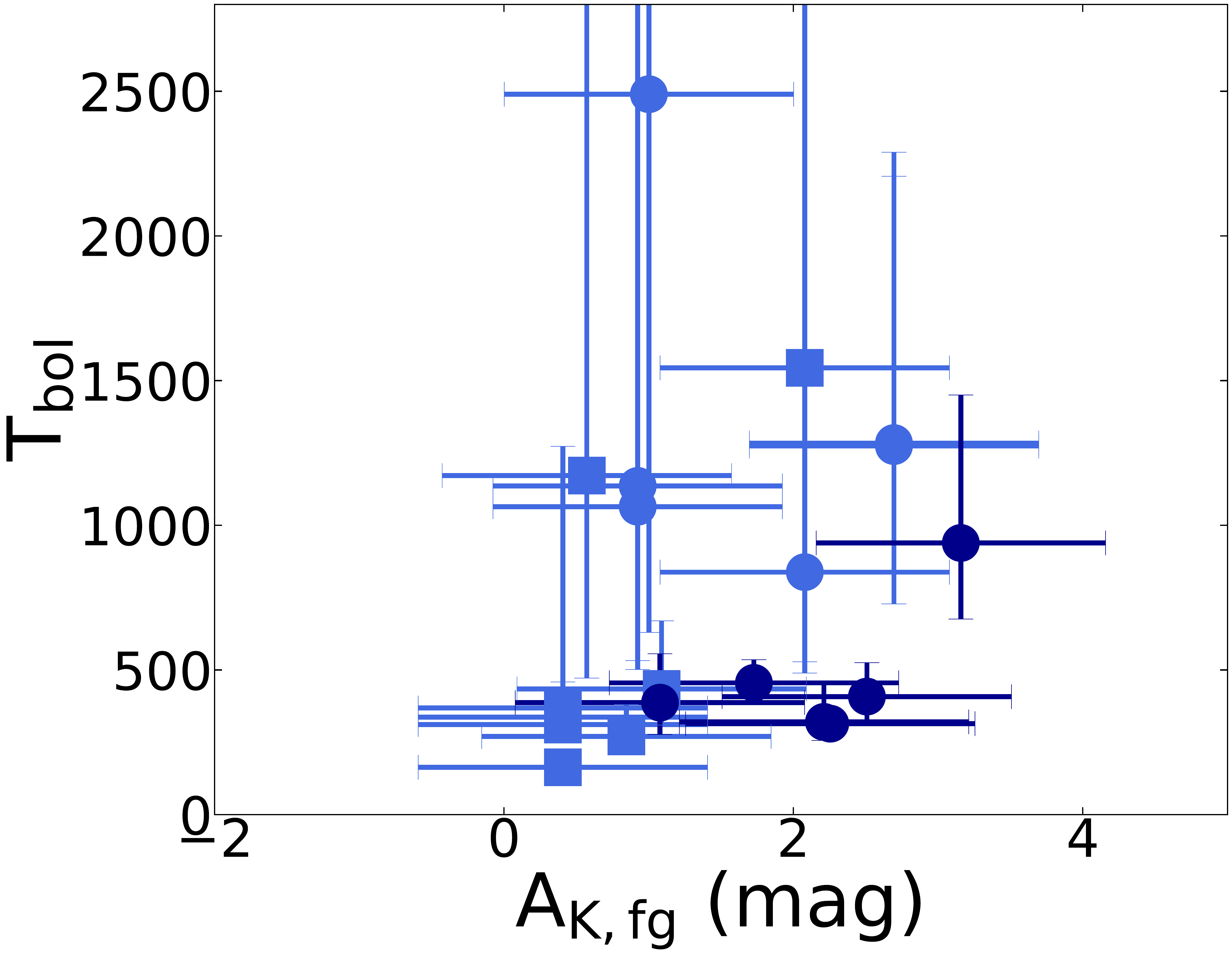} 
    \caption{Bolometric temperature plotted against the computed foreground extinction ($A_{K,fg}$). Symbols are as in Fig.~\ref{fig:f_AKbra_AKpfg}.
    }
    \label{fig:f_Tbol_foreground}
\end{figure}

\begin{figure}
    \centering
    \includegraphics[width=9.0cm]{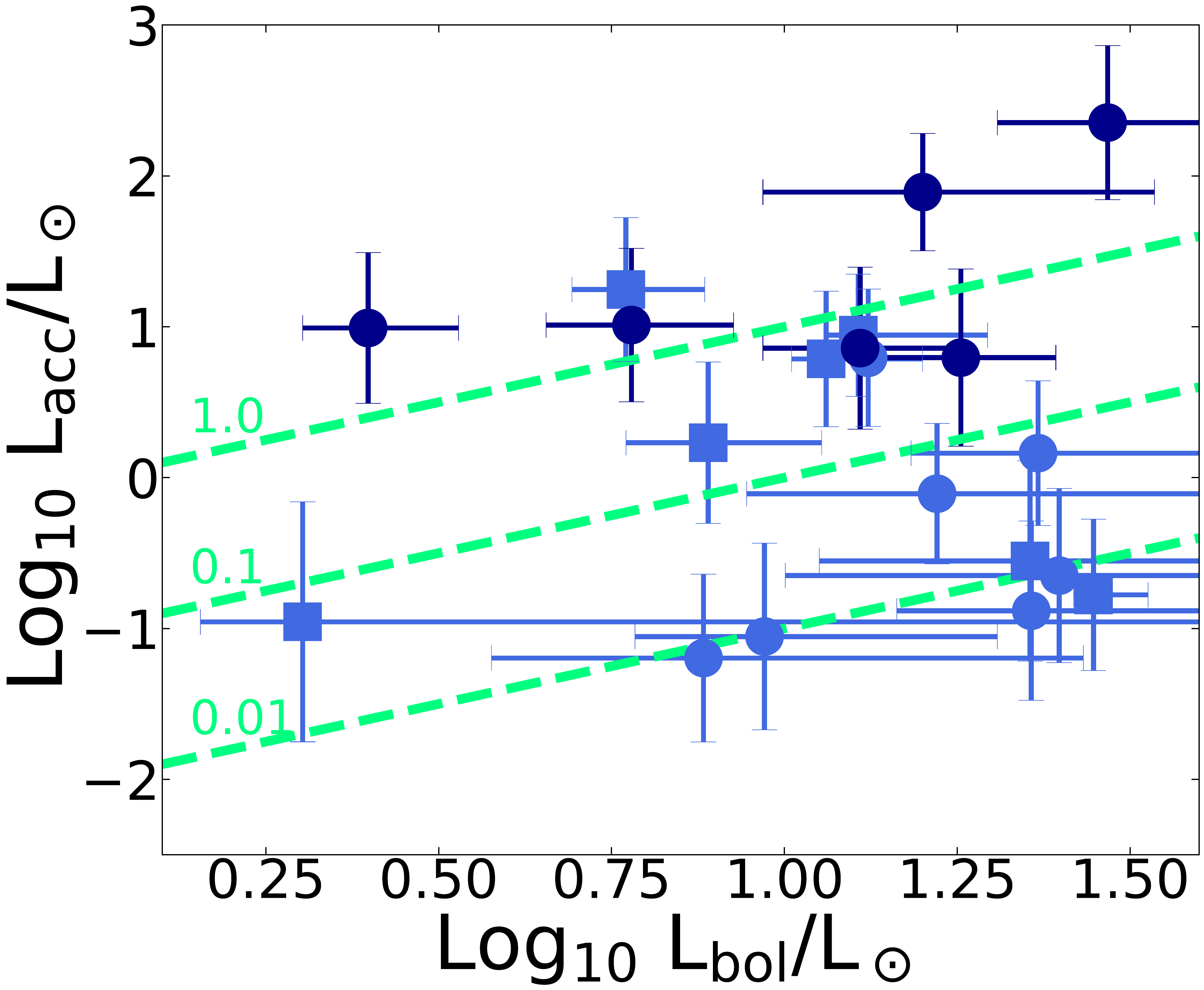} 
    \includegraphics[width=9.0cm]{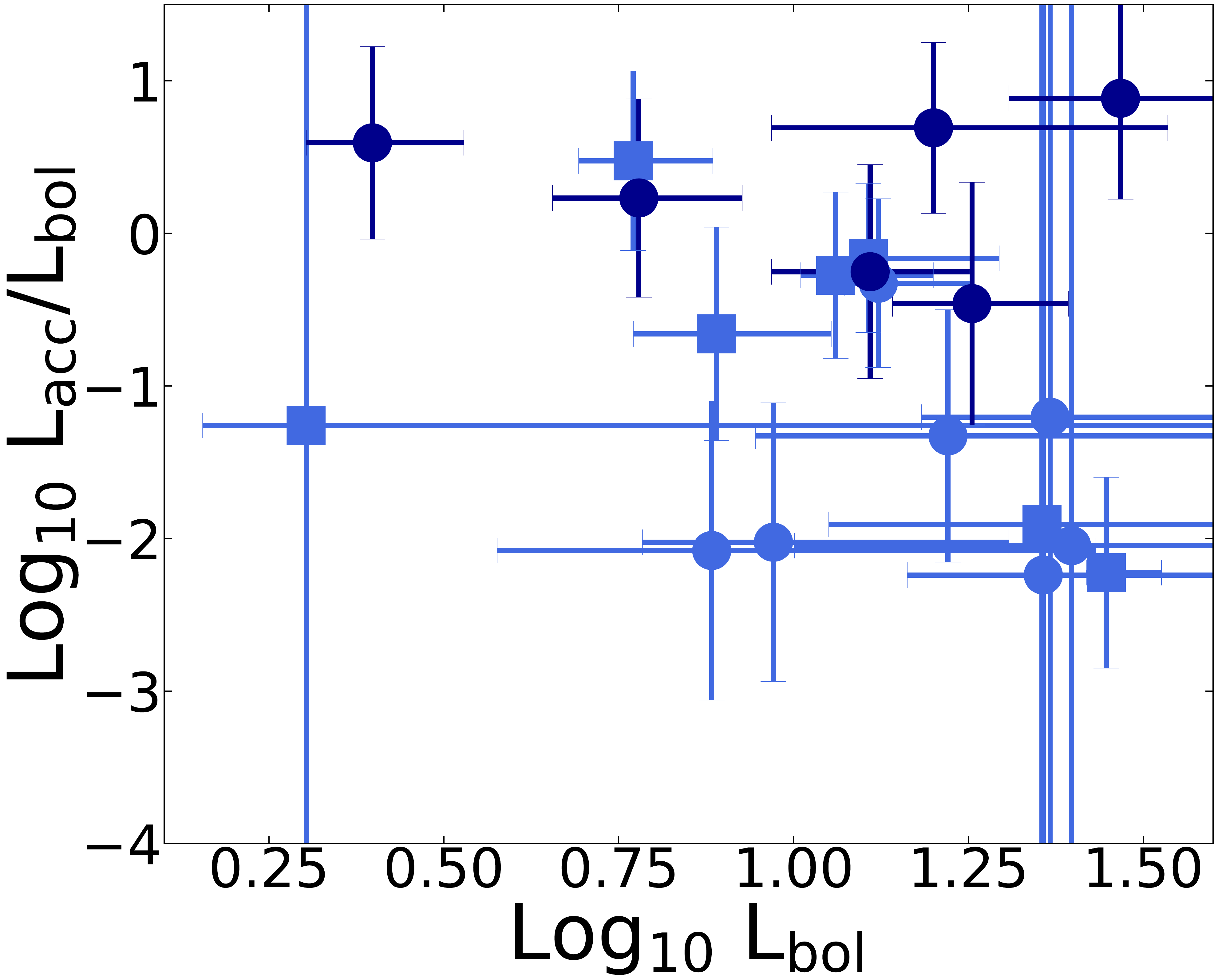} 
    \caption{Accretion luminosity (top) and fraction of accretion luminosity over bolometric luminosity (bottom) as a function of \Lbol. Symbols are as in Fig.~\ref{fig:f_AKbra_AKpfg}. The green dotted lines show the locus of \Lacc/\Lbol=1, 0.1, and 0.01, as marked.}
    \label{fig:f_Lacc_Lbol}
\end{figure}

The uncertainty in A$_{\rm K,fg}$ has a small impact on the derived values of \Lbol\ and \Tbol\ for sources with large values of A$_{\rm K,fg}$ and small values of \Tbol, while the effect is much larger for sources with high values of \Tbol\ (see Fig.~\ref{fig:f_Tbol_foreground}). This is not surprising, as sources with a high \Tbol\ have significant emission at short wavelengths, where an uncertain extinction correction will produce the largest variations. This implies that our computed values of \Lbol\ and \Tbol\ for the youngest sources are less affected by the uncertainty in the estimate of A$_{\rm K,fg}$.
For high \Tbol\ sources, which are mostly part of the Taurus sample, our error bars are generally consistent with these objects being more evolved (possibly Class~II objects previously misclassified as Class I or F).

\begin{figure}
    \centering
    \includegraphics[width=9.0cm]{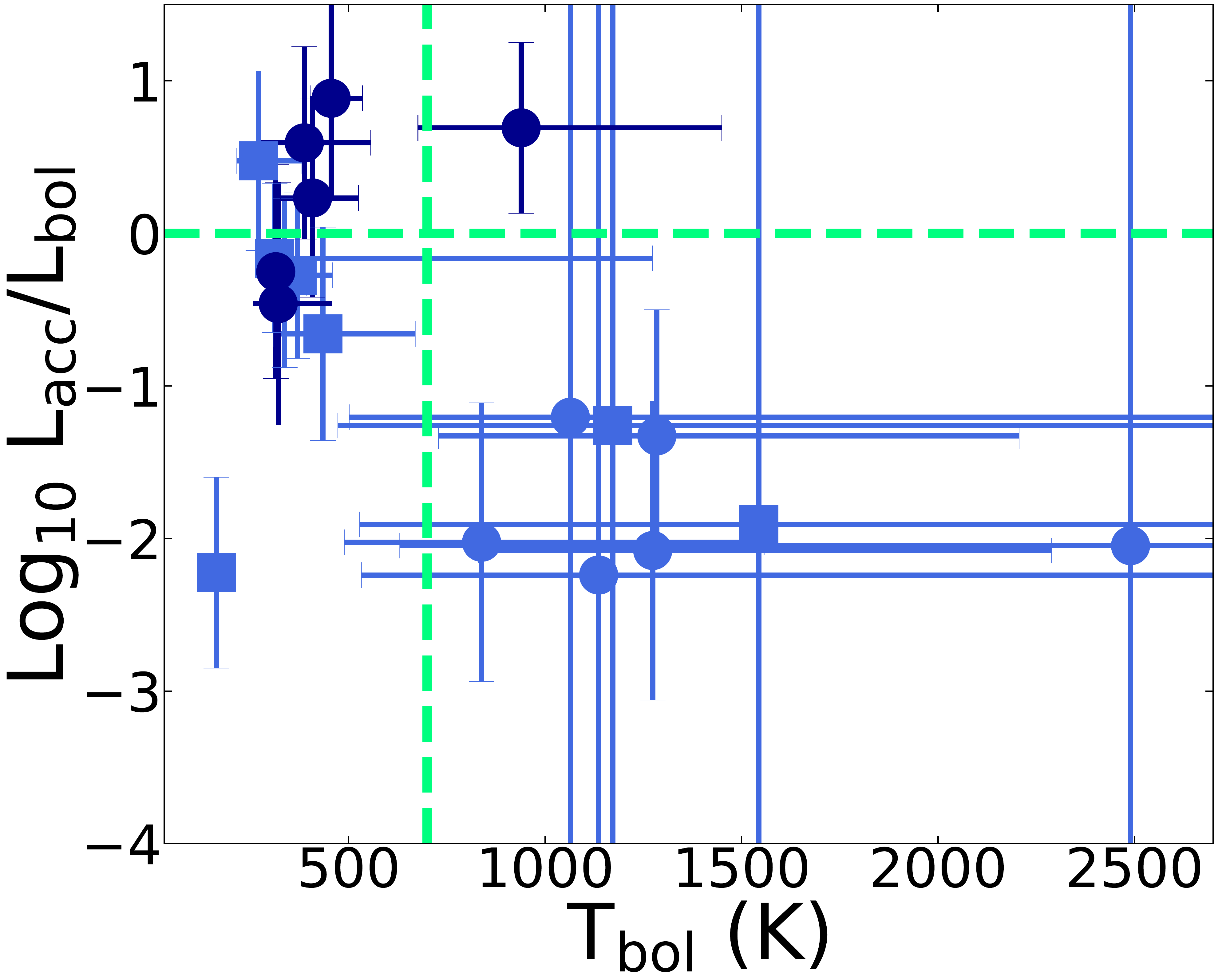} 
    \caption{Ratio of the accretion luminosity to the bolometric luminosity as a function of \Tbol.  Symbols are as in Fig.~\ref{fig:f_AKbra_AKpfg}. The dotted green lines show the \Tbol=700 K and the \Lacc=\Lbol\ locus.}
    \label{fig:f_Lacc-Lbol_Tbol}
\end{figure}

\section {Discussion}
\label{sec:discuss} 

\subsection{Evolution of the \Lacc/\Lbol\ ratio}

The values of \Lacc\ derived in the previous section span a range from $\simless$ 1 to $\sim 100$ \Lsun\ (see Fig.~\ref{fig:f_Lacc_Lbol}). In all but one of the objects with \Tbol$\le 500$~K, within the uncertainties, \Lacc\ is comparable to the total bolometric luminosity, as shown in Fig.~\ref{fig:f_Lacc-Lbol_Tbol}, where the ratio \Lacc/\Lbol\ is plotted as a function of the bolometric temperature \Tbol. Also, all but one of the objects with \Tbol$\ge 800$~K are compatible with \Lacc\ being a small fraction of \Lbol\ (even though the uncertainties in the \Lbol\ measurements are very large for this subsample, as noted above). 
Only one of the low accretors in Taurus has \Tbol $\le 500$~K, and one high accretor in Ophiuchus has \Tbol$\sim 1000$~K. 
{We note that in some cases the derived \Lacc/\Lbol\ ratio is larger than one. This can be due to different reasons, such as the variability of \Lacc\ or \Lbol (see below), or related to the source geometry (see Sect.~\ref{section:geometry}).}

\begin{figure}
    \centering
    \includegraphics[width=9.0cm]{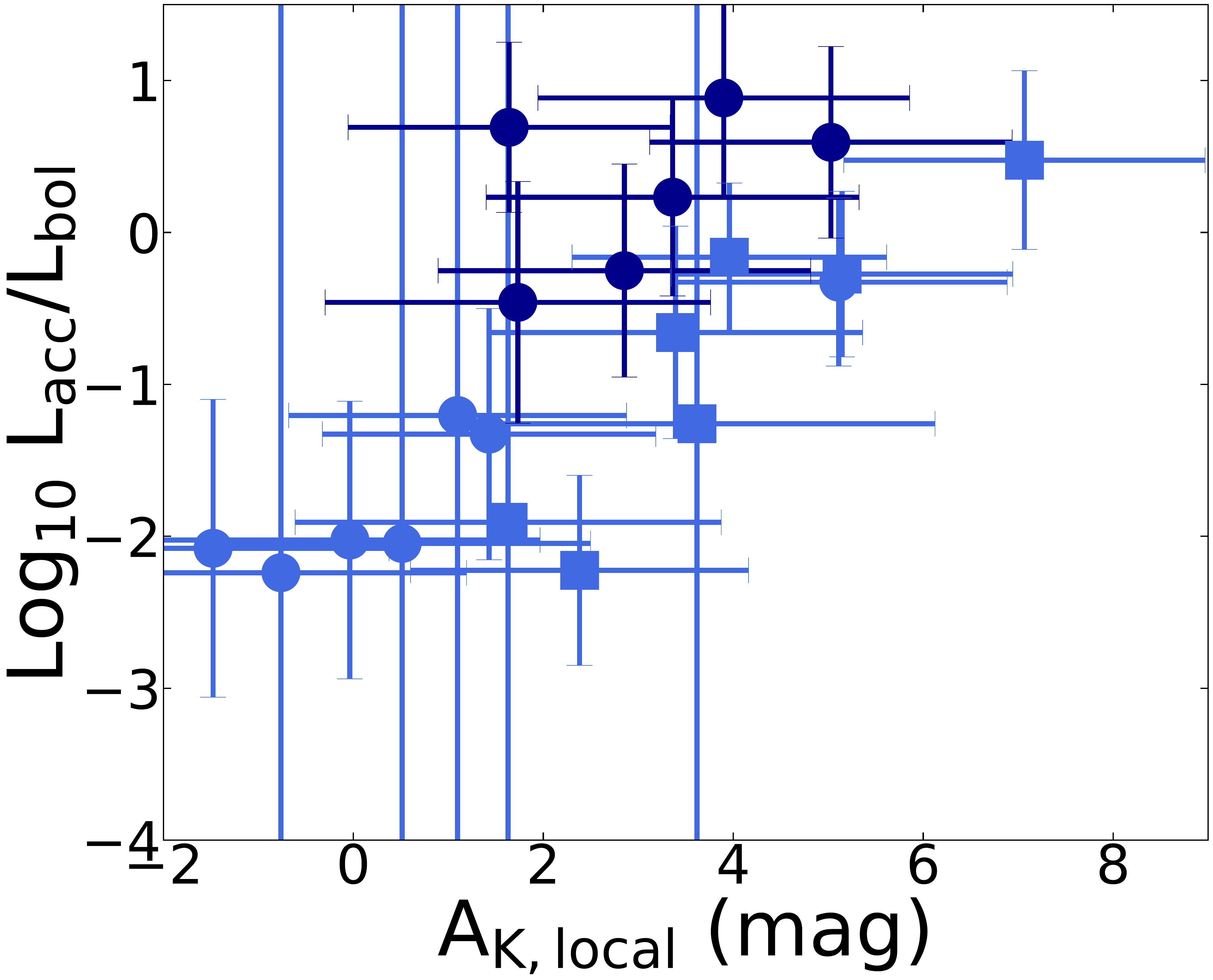} 
    \includegraphics[width=9.0cm]{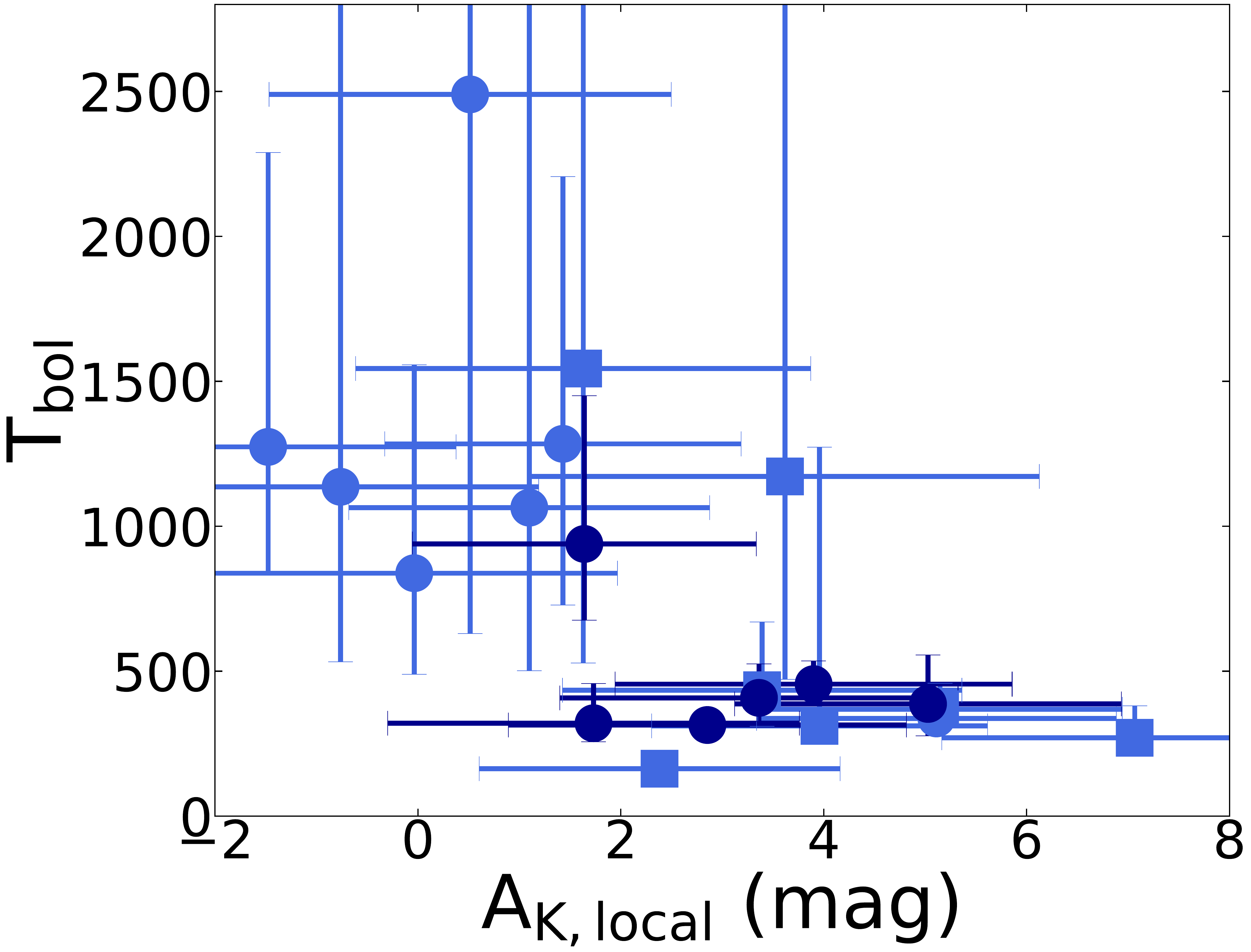} 
    \caption{Top panel: Ratio of \Lacc/\Lbol\  versus the local extinction in the K band \AKlocal. Bottom panel: \Tbol\ versus \AKlocal. Symbols are as in Fig.~\ref{fig:f_AKbra_AKpfg}.}
    \label{fig:f_LaccLbol_AKloc}
\end{figure}

It is generally expected 
that younger objects, where a high fraction of the luminosity is due to accretion, are also more embedded in the parental core and therefore have a higher extinction due to local dust \citep[e.g.,][]{2009ApJS..181..321E}. We computed the local extinction, $\rm A_{local}$, as the difference between the total extinction derived from the line ratios (see Table~\ref{tab:tab_classi}), which depends on the total amount of dust along the line of sight, and the foreground component,  $\rm A_{fg}$. 
Figure~\ref{fig:f_LaccLbol_AKloc} shows the results for the K band. Most low accretors have a low local extinction, ${\rm A_{K,local} \simless 1}$~mag, while most of the high accretors have ${\rm A_{K,local}\simgreat 2}$~mag.
In spite of the large uncertainties that affect some of the results, 
the transition between high to low accretors at \Tbol is $\sim 500-700$K.
This trend is broadly consistent with the expectations from numerical models of star formation. For example, one result of \citet{2024A&A...683A..13L} is that protostars have an initial evolutionary phase in which \Lacc\ dominates the instantaneous total luminosity of the central protostar. Then there is a sharp transition to the total luminosity being dominated by the internal luminosity of the central source. We note that one important difference between the "total" luminosity computed in the models and the "bolometric" luminosity derived from observations is related to the temporal averaging introduced by the reprocessing of radiation in the protostellar envelope. While in observations it is possible to have occurrences of the instantaneous \Lacc\ being larger than the \Lbol, in models the instantaneous total luminosity of the central source is always the sum of the instantaneous accretion and internal luminosities.

The fact that in most of the Ophiuchus Class I sample  the accretion luminosity accounts for all of the \Lbol while  Taurus objects cover a large range of values, from accretion to stellar-dominated luminosity, is likely a selection effect. Indeed,  the Ophiuchus Class I sample contains objects selected because of their  low values of \Tbol\ already corrected for an estimated average foreground extinction \citep{2015ApJS..220...11D}, while the selection of Taurus Class I objects had less stringent criteria. 
The objects classified as Class I purely on the basis of their observed infrared spectral index (uncorrected for foreground extinction) may in fact cover quite a significant range of different evolutionary stages, such as from a time when the central star is still significantly growing to a later phase where the central object has already reached a mass close to the final one, and the photospheric luminosity controls the thermal structure of the surroundings. 

This picture is in agreement with other studies of accretion in young stellar objects. \citet{2023ApJ...944..135F} show that in Class~I sources \Lacc$\simless 0.5$~\Lbol in a sample selected for low values of extinction, which is consistent with relatively evolved Class~I young stellar objects. In their sample, $\sim$80\%\ of the sources show \AK$\le 3$, whereas in our sample (see Fig.~\ref{fig:f_LaccLbol_AKloc}), $\sim$80\%\ of the sources with \AK$\le$3 show \Tbol$\ge$700~K.
\citet{2024ApJ...966...91L} observed Br$\gamma$ line luminosities for a sample of face-on, hence low-extinction, younger Class~0 young stellar objects. They show that at younger ages, the line luminosity is a factor of $\sim$100 higher than in the sample of \citet{2023ApJ...944..135F}.

\subsection{Comparison with numerical simulations}
\label{sec:simul}

Numerical simulations of star formation have now reached a level of sophistication that allows us to estimate the variation of \Lacc\ as the objects evolve in time. To produce a qualitative comparison with our results, we used the \Lacc, L$_{\rm tot}$, age, and mass of the protostars for a fully consistent population formed within a single numerically simulated star-forming cloud in the high resolution non-ideal MHD simulations of \citet{2024A&A...683A..13L}.
We note that for this analysis, the L$_{\rm tot}$ values were computed as the sum of the accretion and intrinsic luminosities of the growing sink particles, which represent the (proto-)stars in the simulation. The value of L$_{\rm tot}$ is thus the total luminosity of the compact protostar, which may differ from the observed \Lbol, as the latter is derived from the line of sight's reprocessed radiation through the envelope (see Sect.~\ref{sec:caveats}).

\begin{figure}
    \centering
    \includegraphics[width=9.0cm]{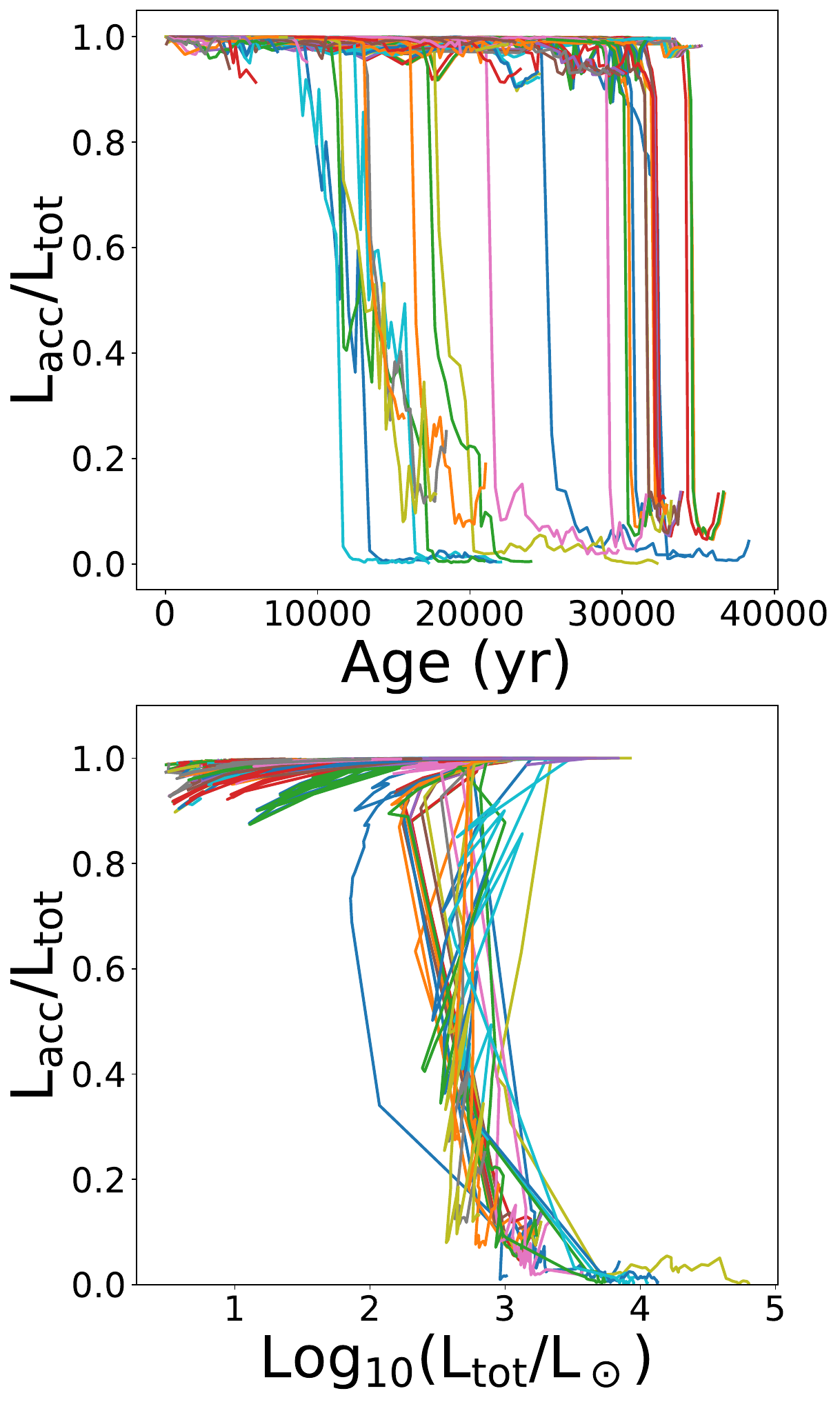} 
    \caption{Ratio of \Lacc/L$_{\rm tot}$\  versus age (top) and L$_{\rm tot}$ (bottom) from the simulations of \citet{2024A&A...683A..13L}. }
    \label{fig:ugo_sink}
\end{figure}

In Figure~\ref{fig:ugo_sink}, we show the values of the ratio \Lacc\ over L$_{\rm tot}$ as predicted by the numerical simulations of \citet{2024A&A...683A..13L}. The simulations show that \Lacc\ dominates  L$_{\rm tot}$ at early stages of evolution, and then the contribution of the accretion to the total luminosity drops very rapidly at later evolutionary stages. In the simulations, the transition occurs when the central protostar reaches a critical internal luminosity threshold. 

The same qualitative behavior is obtained in simulations with different assumptions on the underlying physics \citet[e.g., with mechanical feedback or low magnetization][]{2024A&A...683A..13L}. A more detailed comparison of our measurements with the outcome of currently available numerical simulations remains difficult because of the mismatch in the ages of the objects. 

\section {Caveats}
\label{sec:caveats}

\subsection {Extinction correction}
\label{sec:caveats_extinction}
The results discussed above depend  significantly on the adopted extinction law in the region $\sim 2-4 \mu$m. There is a large uncertainty on what is the most appropriate extinction law for the dense cores of Class~I objects.
When computing the extinction toward the hydrogen recombination lines, we adopted the ratio \ABra/\ABrg=0.54. This is appropriate for the more reddened regions of \roph\ \citep[\AK >1~mag][]{2023Univ....9..364L}, and it is in very good agreement with the interstellar medium extinction curve of \citet{2001ApJ...548..296W} for $\rm{R_V}$=5.5. The values of \AK\ obtained for our Class I objects (Table~\ref{tab:tab_classi}) mostly  fall in this range, suggesting that, at least,  there is no inconsistency in our results. 

Theoretically, grain growth from the interstellar medium size is expected to occur in cores, reaching maximum grain sizes that could reach a few tens of microns for a density of $10^6$ cm$^{-3}$ and age of 1~Myr \citep{1994AA...291..943O,2009AA...502..845O,2011AA...532A..43O,2023MNRAS.518.3326L}. In these conditions, \citet{1994AA...291..943O} found \ABra/\ABrg\ values of $\sim 0.55$ for ice-coated grains from the standard size distribution \citep{1977ApJ...217..425M}, similar to the value 0.54 adopted in this paper.  The authors of \citet{2011AA...532A..43O} followed the evolution of grain properties in time and found values  increasing from $\sim 0.3$ to $\sim 0.7$ at 1 Myr. Observationally, recent results by \citet{2024ApJ...965...29L} on four cores (two starless, one Class~0, and one Class~I) show a flat extinction law very similar among the four cores studied that results in a \ABra/\ABrg\ ratio of $\sim 0.7$. Interestingly, this is similar to the most recent values for  interstellar medium extinction \citep{2020ApJ...895...38H}. However, there is quite a spread in the observational results \citep[see, e.g.,][]{2024ApJ...965...29L}. 

A flatter extinction curve results in a larger value of the extinction than we derived from the observed line flux ratios. This can have a significant effect, as it increases the line luminosities by a large factor and therefore also the accretion luminosity. For example, when adopting the \citet{2024ApJ...965...29L} extinction curve, the values of \ABra\ in Table~\ref{tab:tab_classi} roughly double, and \LBra\ (hence, \Lacc) increases from a factor of a few for \ABra$\sim$1--2 mag to a factor of $\sim 50$ for objects with \ABra$\sim$4 mag.
Such very high values of \Lacc\ would be difficult to reconcile with the observed bolometric luminosities.

\subsection {Geometry of the source and scattering}
\label{section:geometry}

What we discussed in the previous paragraph is the uncertainty on the ``pencil beam'' extinction due to dust properties. When observing an unresolved spatially extended source, the unresolved geometry plays an important role in the ``effective'' beam-averaged extinction. The effect of source geometry has been investigated through simulations of the emerging spectral energy distribution of protostars with complex geometry \citep[e.g.,][]{1993ApJ...414..676K,1997ApJ...485..703W,2003ApJ...591.1049W}. In particular, \citet{2003ApJ...591.1049W} have discussed the effects of the geometry and inclination angle of Class~I young stellar objects on the estimates of the source bolometric luminosity and infrared colors.

The general conclusion is that the effects on \Lbol\ are small; the variations of the estimate of \Lbol\ for the full range of inclinations is found to be smaller than a factor of two.
This is a small uncertainty for our estimates of \Lbol given all the other sources of errors. On the other hand, the effect of geometry on the infrared colors is more relevant. The results of \citet{2003ApJ...591.1049W} demonstrate that there could be a significant systematic uncertainty in the use of infrared colors to derive extinction due to the unknown viewing angle, especially for sources with the disk axis close to the plane of the sky. {These effects have been confirmed observationally by a spectral energy distribution fitting of large samples of young stellar object photometry \citep{2016ApJS..224....5F,2023ApJS..266...32P}. The total model luminosities can exceed the measured \Lbol\ by a factor of up to three for the most inclined systems.}
These uncertainties can only be solved by the combination of more detailed observations, in order to derive the source geometry, and the dust and line radiation transfer through the disk-envelope system. 
We note that these uncertainties affect all attempts to measure accretion luminosity using line emission in protostars.

\section {Class II as templates for Class I accretion properties}
\label{sec:Discussion_ClassII}

{The use of hydrogen line ratios to measure the extinction toward Class I objects in a manner similar to what we have done here is not new. It has been used, for example, by \citet{2007AJ....133.1673B},  \citet{2012A&A...538A..64C}, and 
\citet[]{2013ApJ...778..148E}.
A correct choice of the intrinsic line ratios is particularly important in the case of Class I.
\citet{2007AJ....133.1673B} also used the \Bra/\Brg flux ratio to compute the extinction. The main difference with our approach is that they adopted an intrinsic ratio of approximately three, which is the theoretical value expected from Case B recombination lines \citep{1987MNRAS.224..801H}, rather than the value measured in Class II ($\sim 1$). As a consequence, the extinction derived by \citet{2007AJ....133.1673B} is significantly lower than our determinations, and  \Lacc\ is smaller by a factor that ranges from roughly one for the less reddened objects to $\sim 40$ for the most reddened ones.

Our choice of using the observed Class II line ratios is new and potentially  important. In the case of the line ratios used in this paper
(see Sec.\ref{sec:results_classII}), the derivation of the extinction and accretion luminosity of the Class I is made easy by our finding that the two ratios, \Bra/\Brg\ and \Pfg/\Brg, in Class II
are independent of stellar and accretion properties (see Fig~\ref{fig:f_ratio_ClassII}) so that a mean value can be derived with an uncertainty of only $\sim$ 20\%. We think that this is probably the case for many other hydrogen line ratios, as the studies of hydrogen line luminosities in Class II have found that the luminosity of each individual line correlates with \Lacc\ almost linearly \citep{2017A&A...600A..20A,2017MNRAS.464.4721F,2024AJ....167..232T}. }

One caveat, already mentioned in Sec.~\ref{sec:results_classII}, is that the lack of low-mass Class II stars accreting at very high rates forces us to use intermediate-mass Class II as a proxy for Class I objects. At present, most of the available evidence indicates that in Class II, the total accretion luminosity is distributed roughly in the same proportion among the many emission lines observed in the spectra independently of such properties as age and mass. However, this is a crucial aspect of the procedure outlined in this paper, and it should be further analyzed. 

The results of this paper show that it is necessary to have measurements of multiple line ratios extending from the near-infrared to the mid-infrared, possibly observed simultaneously, for a large number of objects so that accurate mean values can be computed and suitable line ratios are selected. Such a dataset will be provided by JWST spectroscopy of Class II objects, and it could provide  the base for determination of the accretion luminosity in embedded objects, following the classic method outlined in this paper. Moreover, if multiple mean values of line ratios spread over a large wavelength interval are available, one could also obtain very valuable constraints on the extinction law itself \citep[see, e.g.,][]{2024A&A...688A.111R}.

\section{Summary and conclusions}
\label{sec:concl}

In this paper, we have presented simultaneous \Brg\ ($\lambda=2.17 \mu$m), \Pfg\ ($\lambda=3.74 \mu$m), and \Bra\ ($\lambda=4.05 \mu$m) observations with the SpeX instrument of a sample of highly accreting Class~II stars and Class~I protostars. The aim of this study was to characterize the line emission properties of Class~II stars and use these as a template to derive accurate accretion luminosities for the Class~I protostars. Our findings our outlined as follows: 

\begin{itemize}
    \item From the analysis of the three lines in our Class II sample, we find that their flux  ratios show a very low dispersion with no dependence on the photospheric or accretion parameters.  
    {We derived mean values of the two ratios, \Bra/\Brg\ and \Pfg/\Brg, and
    computed the correlations between the accretion luminosity and the line luminosities of \Bra\ and \Pfg (see Equations~\ref{eq:lacc-lbra} and~\ref{eq:lacc-lpfg}).}
    \item Under the commonly used assumption that the accretion engine inside Class~I protostars is similar to more evolved Class~II stars, we developed a new method to derive a measurement of the extinction that affects the line emitting region by comparing the observed line ratios in Class~I with the mean values measured in Class~II. 
    \item This method can be extended to more line ratios and to longer wavelengths using, for example, JWST. Doing so would allow for the possibility of constraining the infrared extinction law in protostars and hence the properties of dust grains.
    \item We applied our method to a sample of young stellar objects in different evolutionary stages in the Taurus and Ophiuchus star-forming regions. Our limited sample shows that accretion luminosity either fully dominates the bolometric luminosity of the young stellar objects or is negligible (as in Class~II). This transition is correlated with the evolutionary stage, as traced by \Tbol. We have shown that this behavior is qualitatively consistent with the prediction of numerical simulations.
\end{itemize}

We have discussed several difficulties in interpreting observations of Class~I protostars that affect our and other methodologies presented in the literature. Nevertheless, our results are intriguing, and the new methodology has a great potential to be applied to statistically significant samples of protostars.  

\begin{acknowledgements}
We used the remote observing feature of the Infrared Telescope Facility, which is operated by the University of Hawaii under contract 80HQTR24DA010 with the National Aeronautics and Space Administration. We thank the kind and effective support of the telescope operators on Mauna Kea.
This work was partly supported by the Italian Ministero dell Istruzione, Universit\`a e Ricerca through the grant Progetti Premiali 2012 – iALMA (CUP C$52$I$13000140001$).
This project has received funding from the European Union's Horizon 2020 research and innovation programme under the Marie Sklodowska-Curie grant agreement No 823823 (DUSTBUSTERS),  from the European Research Council (ERC) via the ERC Synergy Grant {\em ECOGAL} (grant 855130), and the ERC Starting Grant {\em WANDA} (grant 101039452). Views and opinions expressed are however those of the author(s) only and do not necessarily reflect those of the European Union or the European Research Council Executive Agency. Neither the European Union nor the granting authority can be held responsible for them.
RSK also acknowledges financial support from the German Excellence Strategy via the Heidelberg Cluster ``STRUCTURES'' (EXC 2181 - 390900948), and from the German Ministry for Economic Affairs and Climate Action in project ``MAINN'' (funding ID 50OO2206); in addition, he thanks the 2024/25 Class of Radcliffe Fellows for highly interesting and stimulating discussions.
\end{acknowledgements}


\bibliographystyle{aa} 
\bibliography{aa54149-25} 


%

\begin{appendix} 

\section{Line spectra}
\label{sec:app_spectra}

\begin{figure*}[!hb]
    \centerline{
      \includegraphics[width=8.cm]{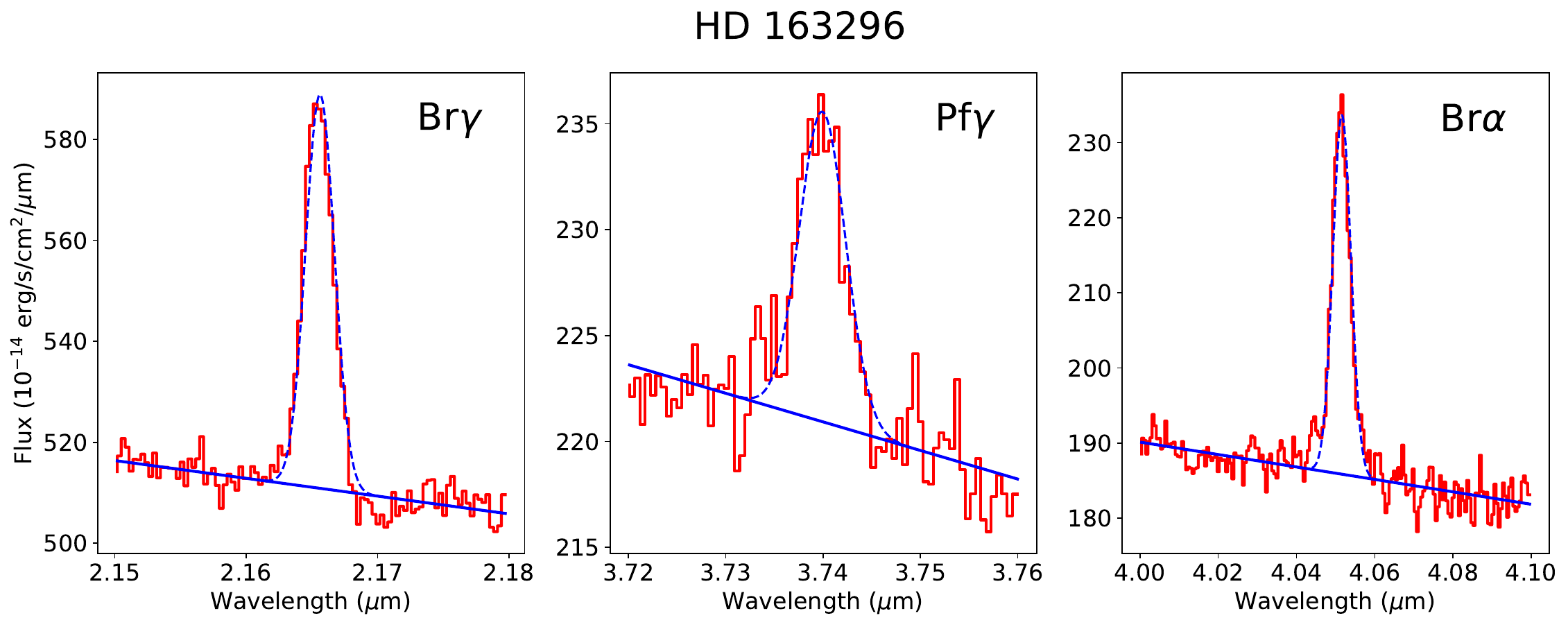} 
      \,\,\,\,
      \includegraphics[width=8.cm]{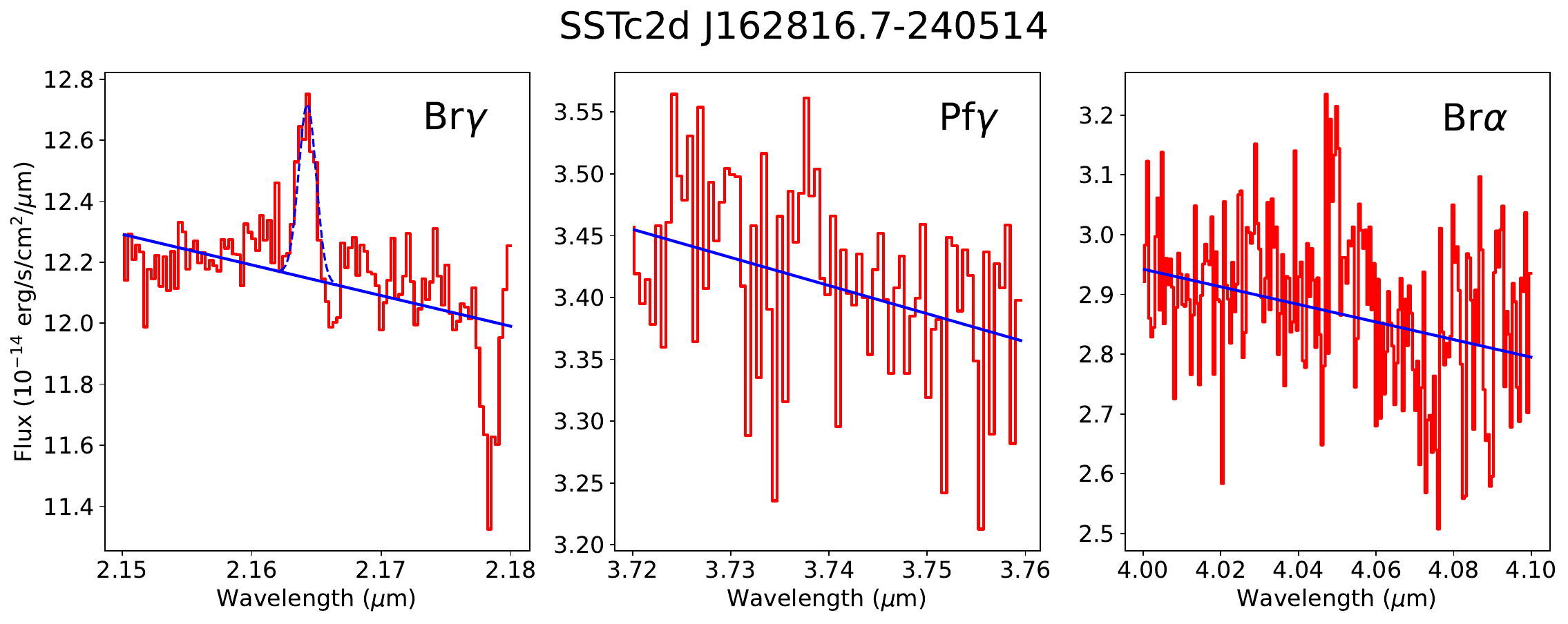} 
    }
    \centerline{
      \includegraphics[width=8.cm]{f_lines_SSTc2dJ162816.7-240514.pdf} 
      \,\,\,\,
      \includegraphics[width=8.cm]{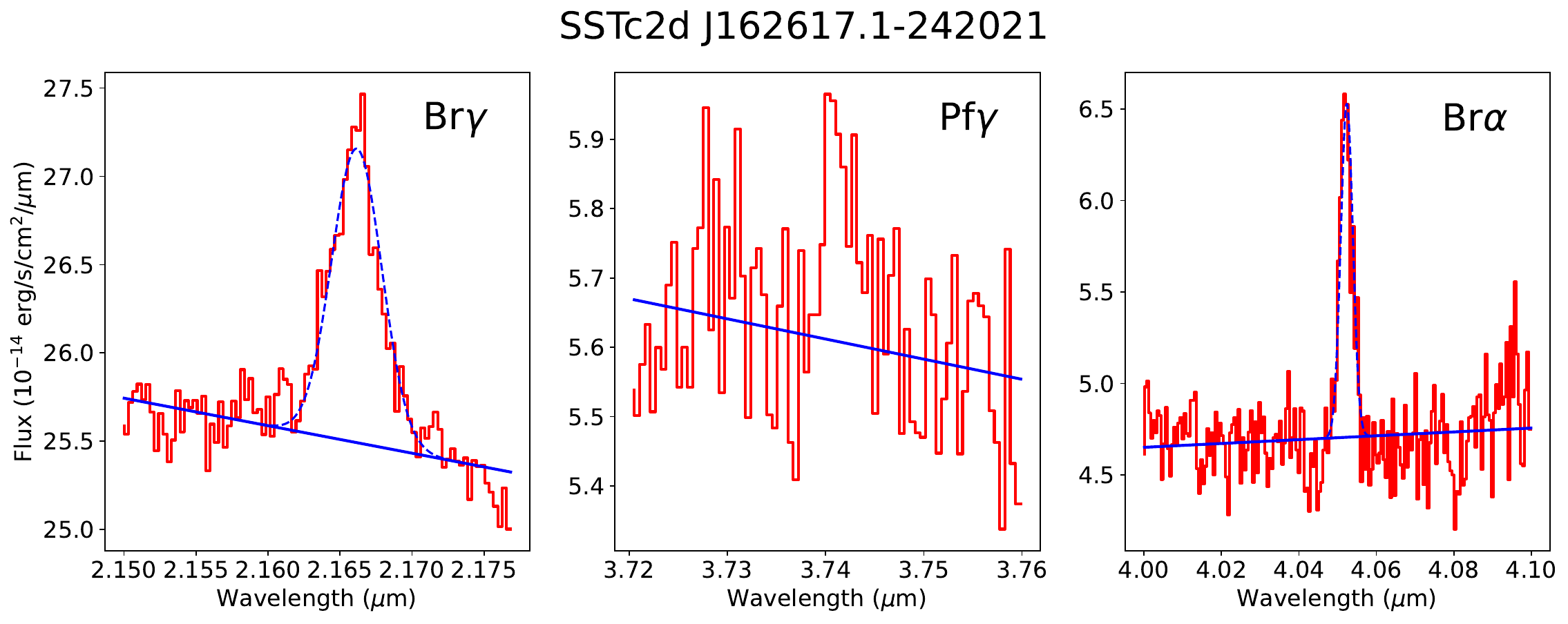} 
    }
    \centerline{
      \includegraphics[width=8.cm]{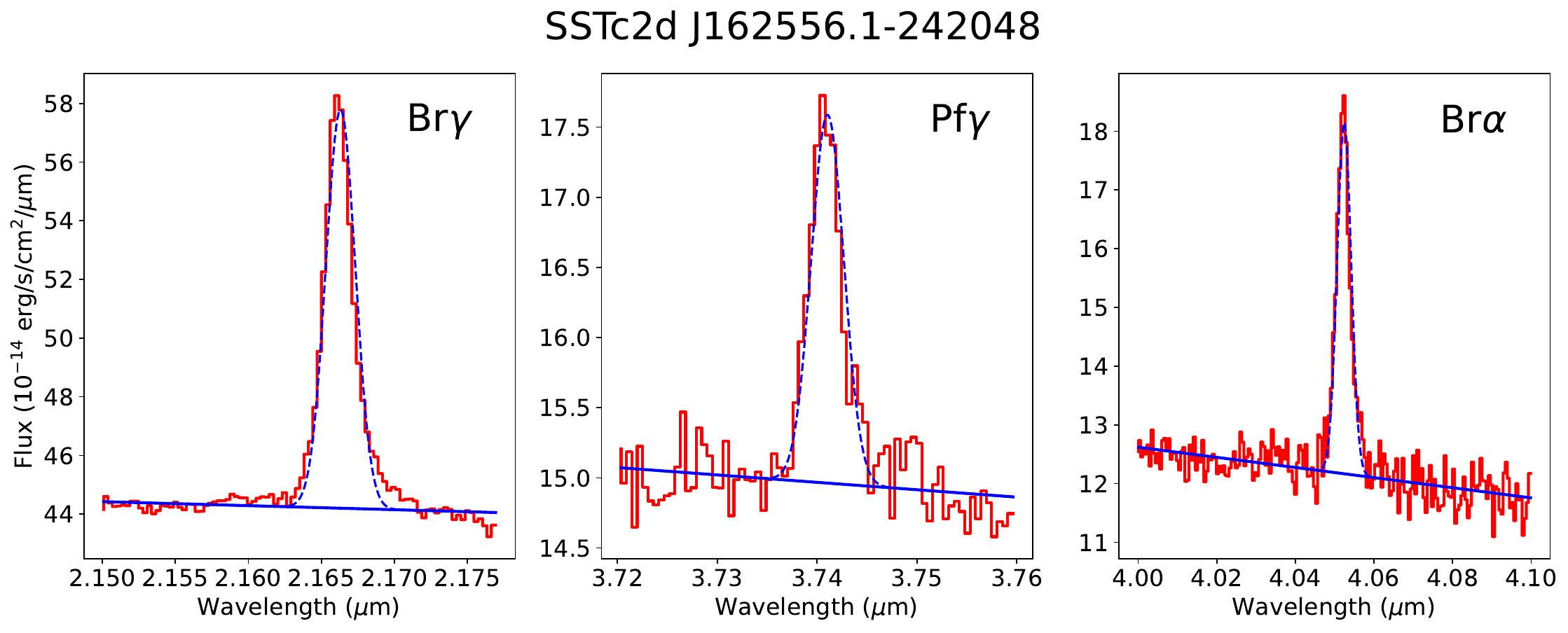} 
      \,\,\,\,
      \includegraphics[width=8.cm]{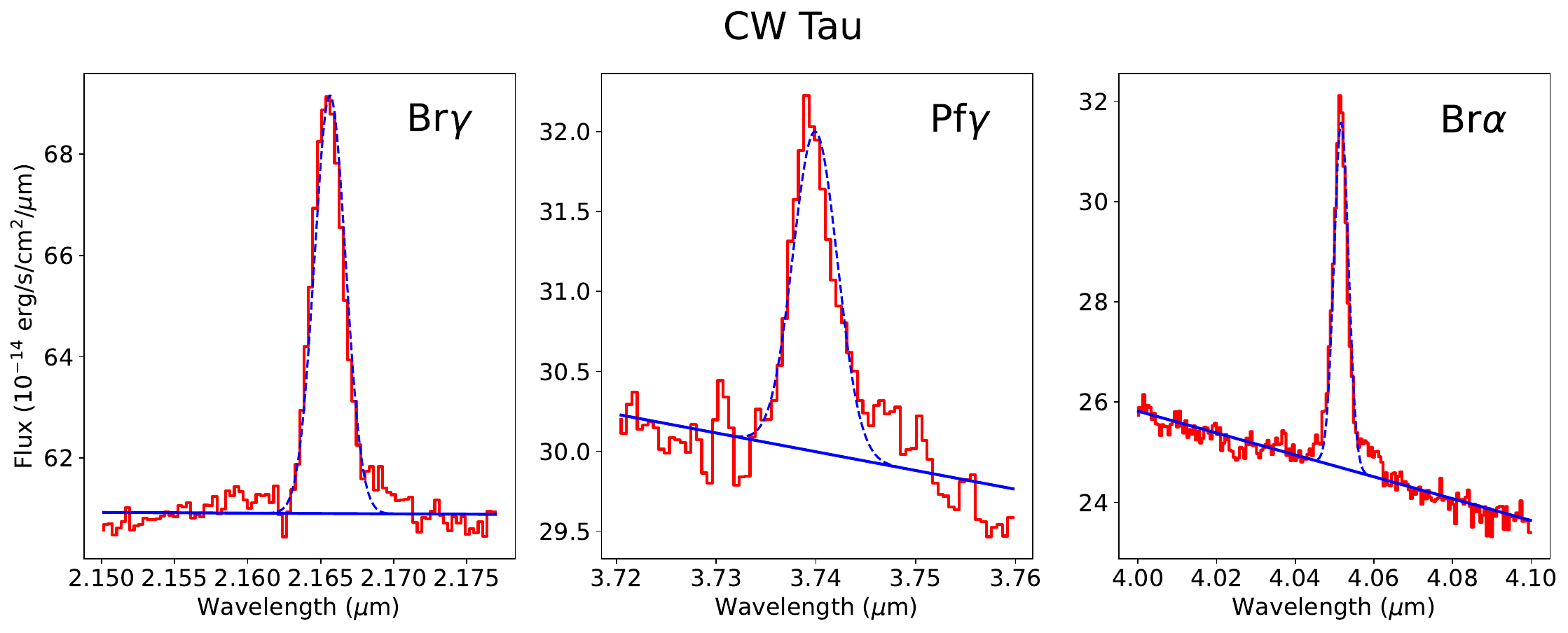} 
    }
    \centerline{
      \includegraphics[width=8.cm]{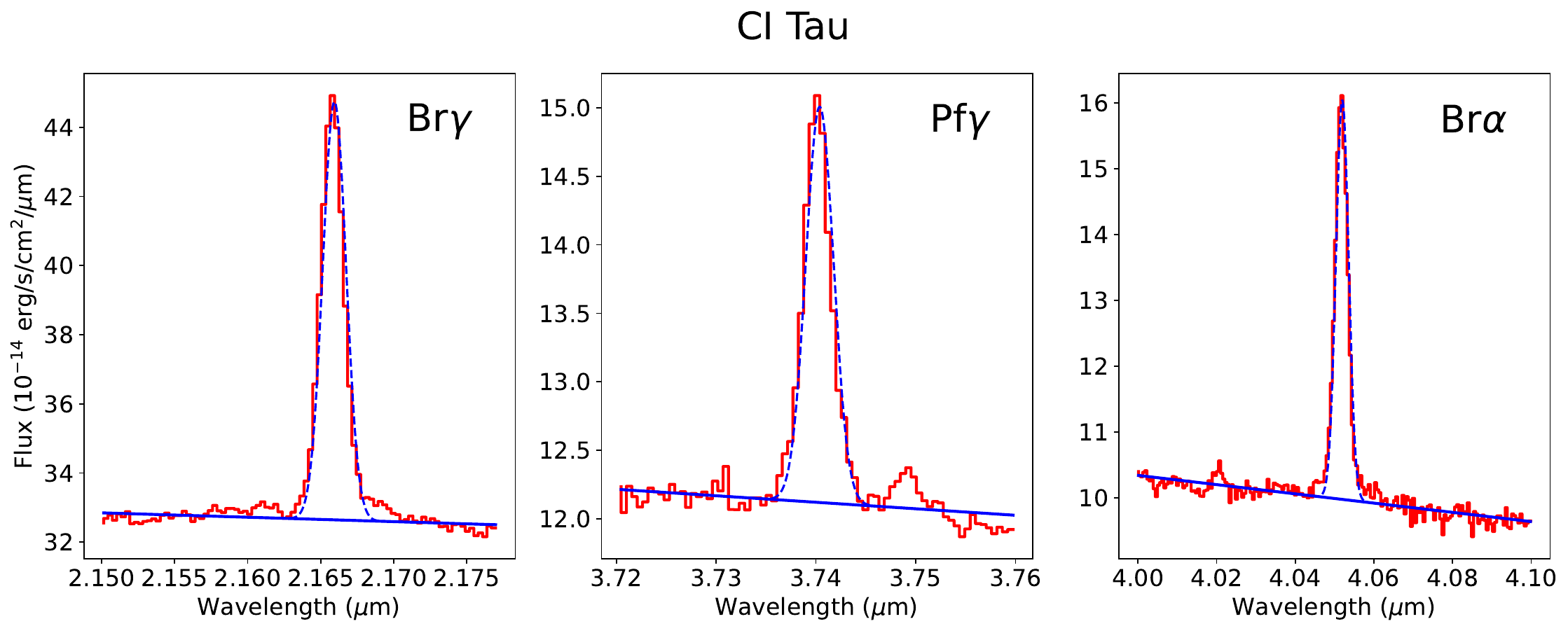} 
      \,\,\,\,
      \includegraphics[width=8.cm]{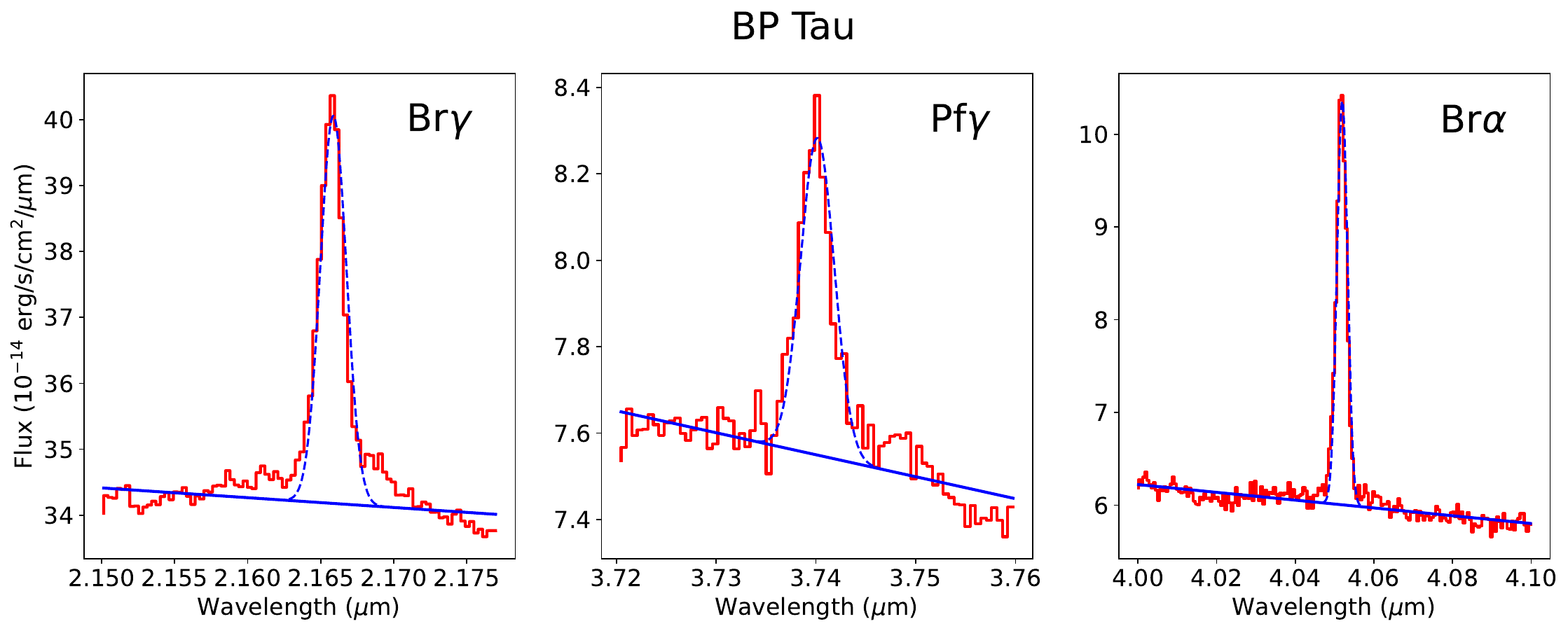} 
    }
    \caption{Spectra of \Brg, \Pfg, and \Bra\ (red line) for the Class~II sources observed with SpeX. The fitted continuum is shown as a blue solid line. When the line is considered a detection, the Gaussian fit used to estimate the flux is shown as a dashed line.}
    \label{fig:f_class_ii_spectra}
\end{figure*}

In this Appendix we report the observed spectra in the region of the \Brg, \Pfg, and \Bra\ lines.
In each panel we show the spectrum in red, the fitted continuum as a solid blue line, and the gaussian fit used to estimate the line flux as a dashed blue line (only for the sources where the line was detected).
Figure~\ref{fig:f_class_ii_spectra} shows the spectra for Class~II sources, while Figs.~\ref{fig:f_class_i_spectra_a} and~\ref{fig:f_class_i_spectra_b} show the Class~I spectra.

\begin{figure*}[!ht]
    \centerline{
      \includegraphics[width=8.8cm]{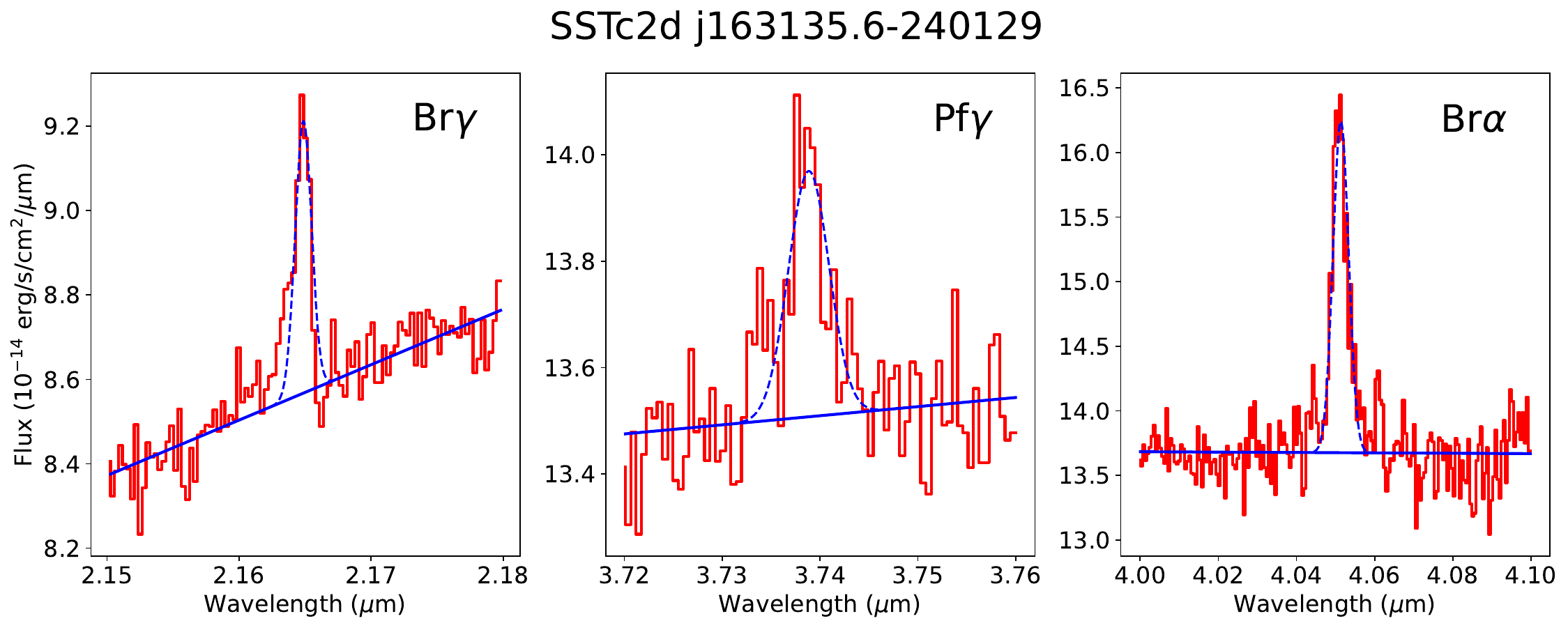} 
      \,\,\,\,
      \includegraphics[width=8.8cm]{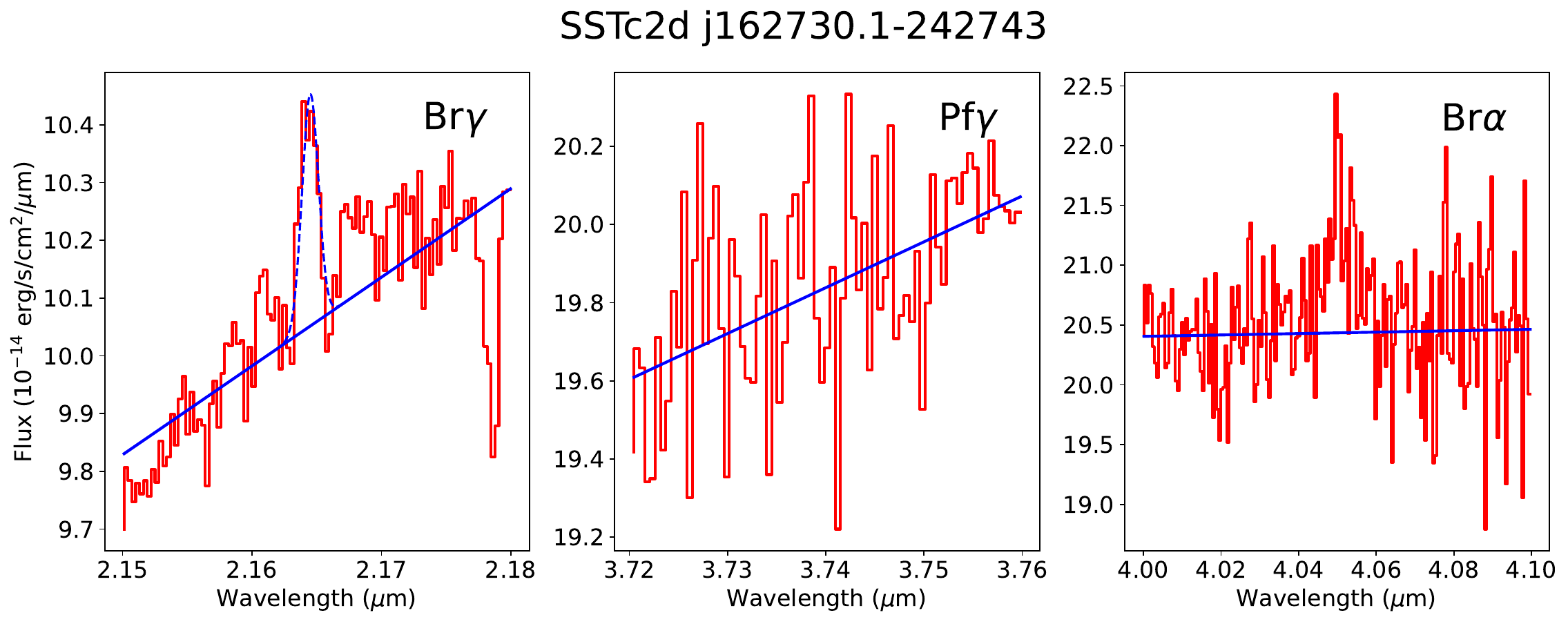} 
    }
    \centerline{
      \includegraphics[width=9.cm]{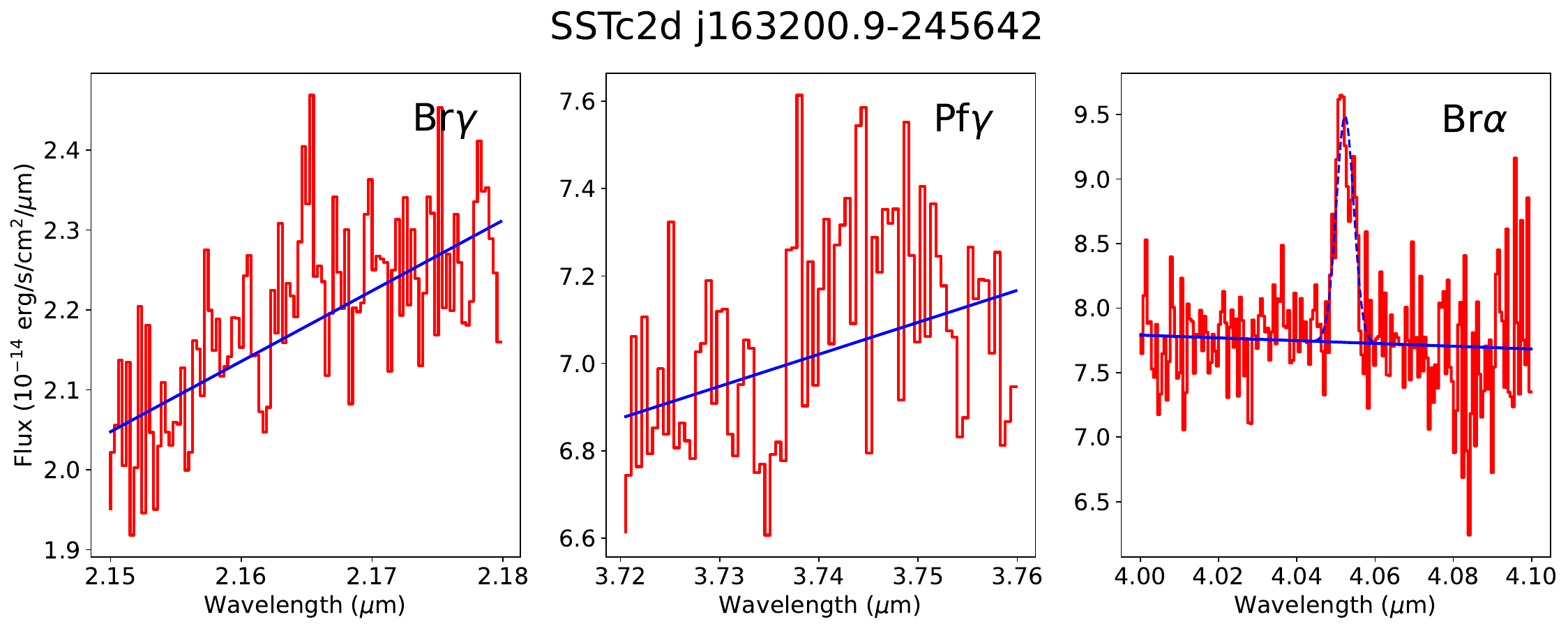} 
      \,\,\,\,
      \includegraphics[width=8.8cm]{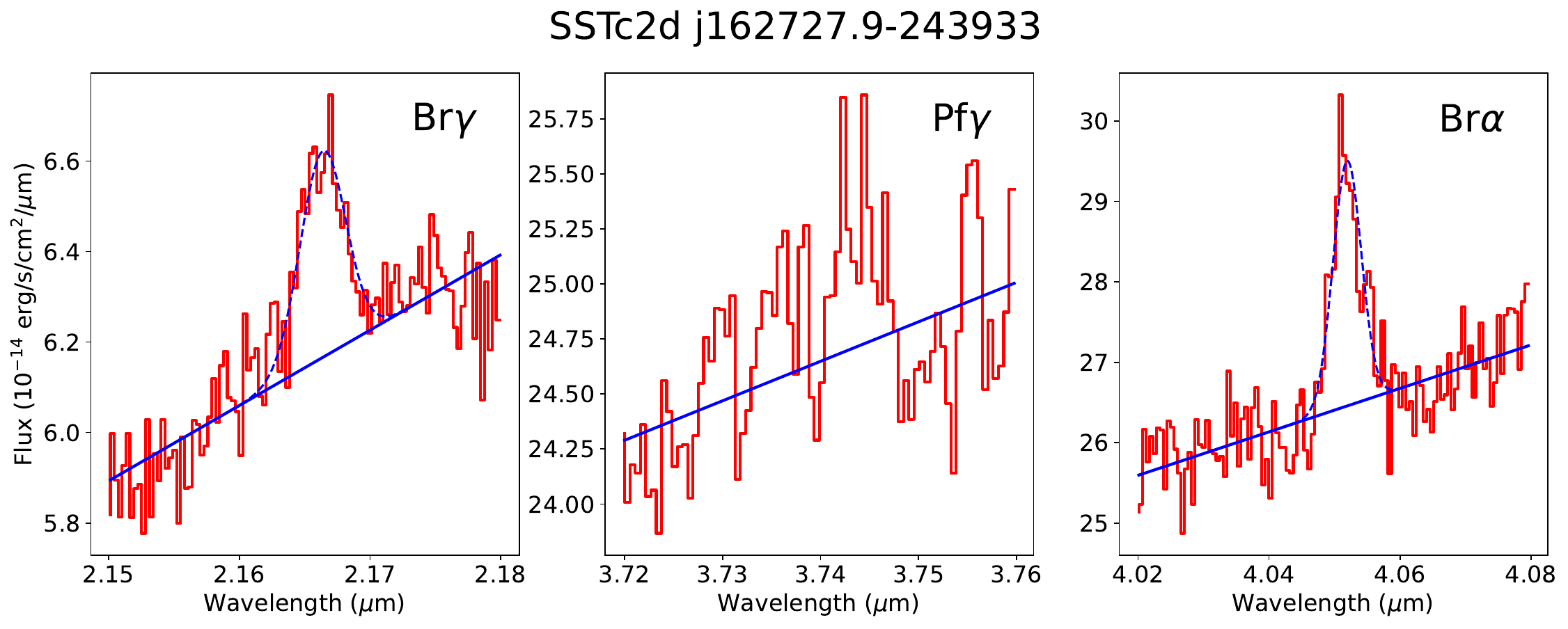} 
    }
    \centerline{
      \includegraphics[width=9.cm]{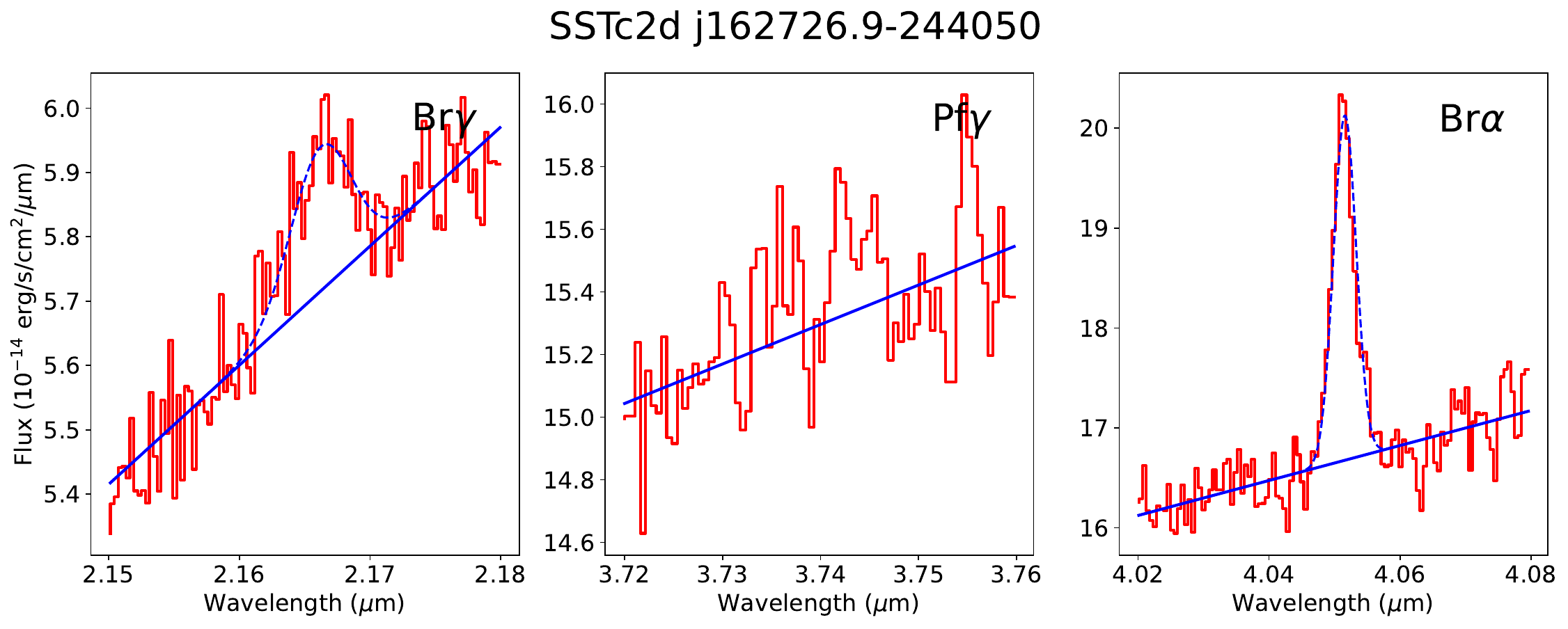} 
      \,\,\,\,
      \includegraphics[width=8.8cm]{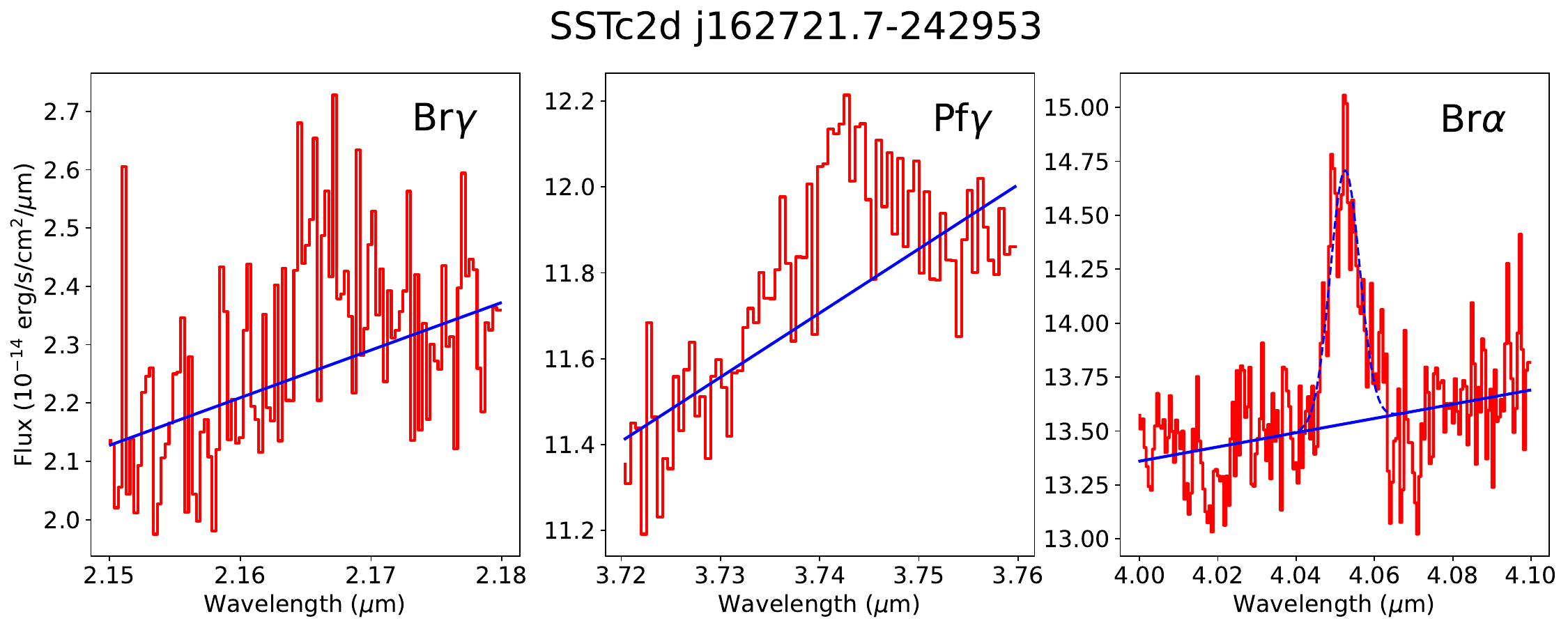} 
    }
    \centerline{
      \includegraphics[width=9.cm]{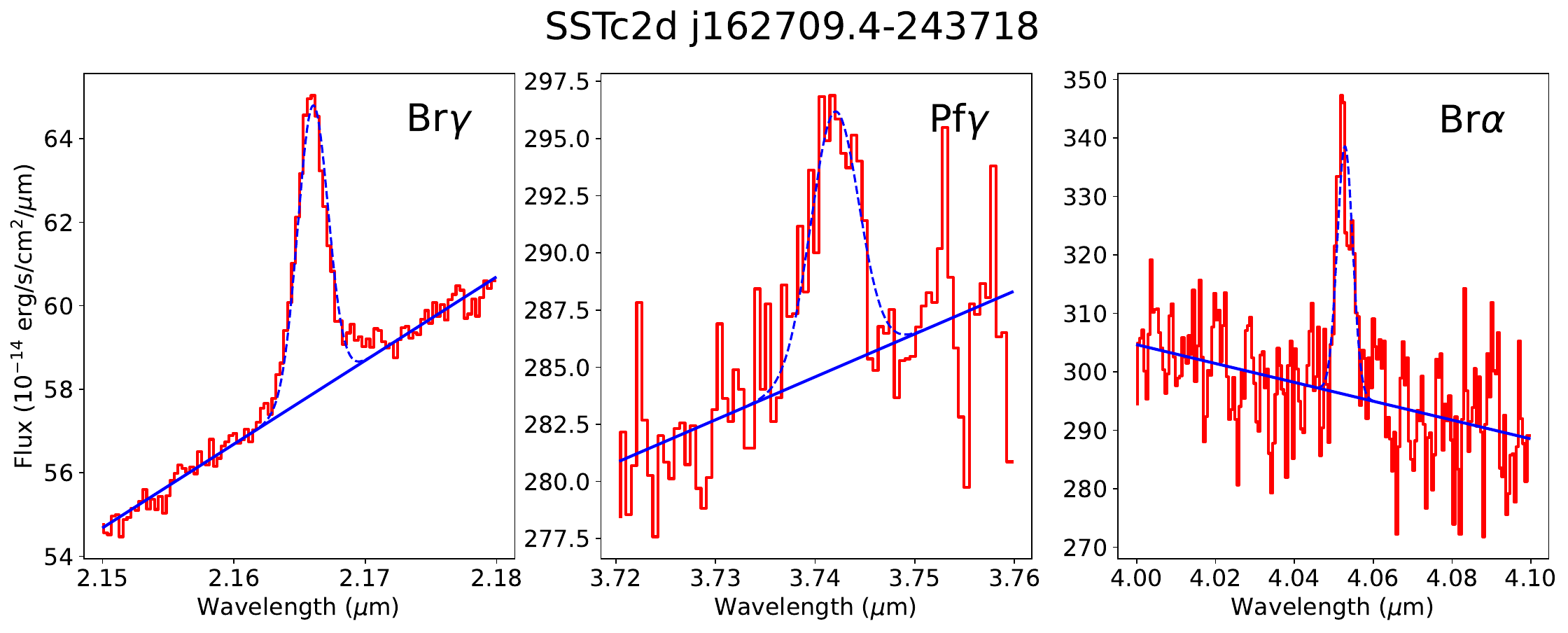} 
      \,\,\,\,
      \includegraphics[width=8.8cm]{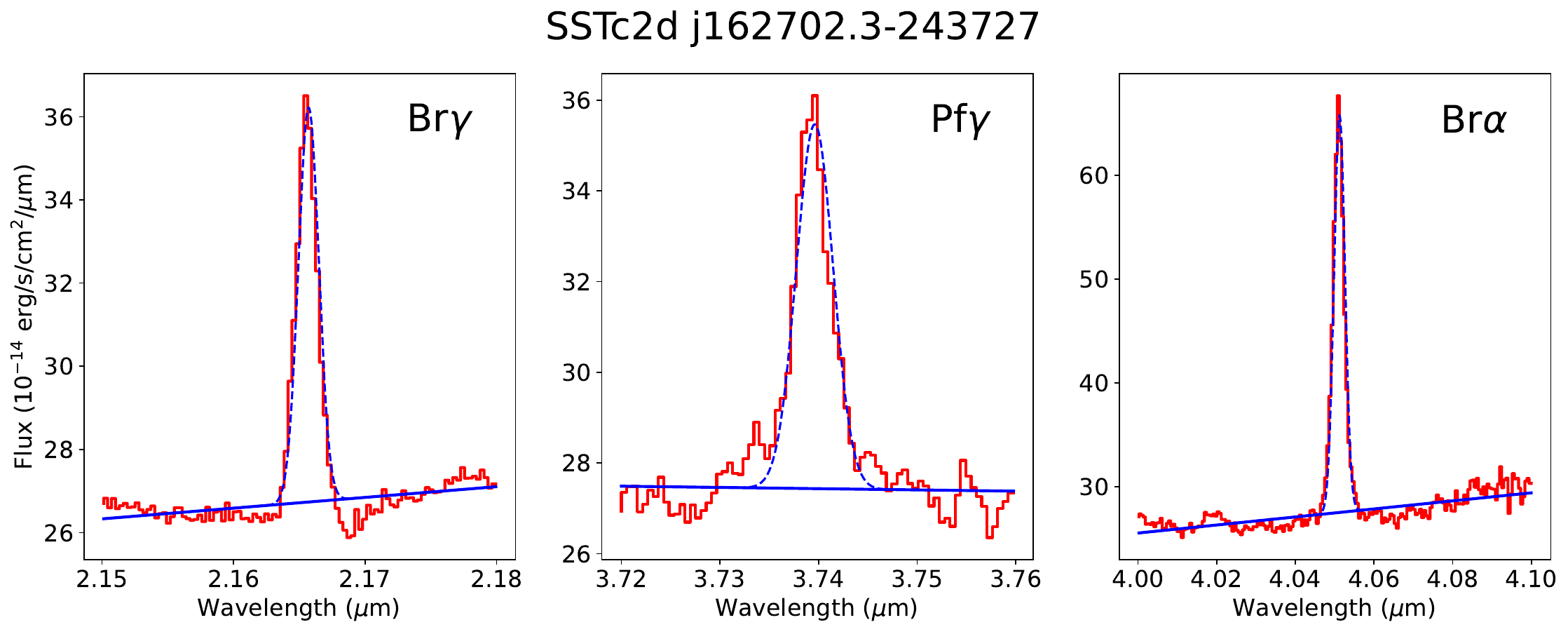} 
    }
    \centerline{
      \includegraphics[width=9.cm]{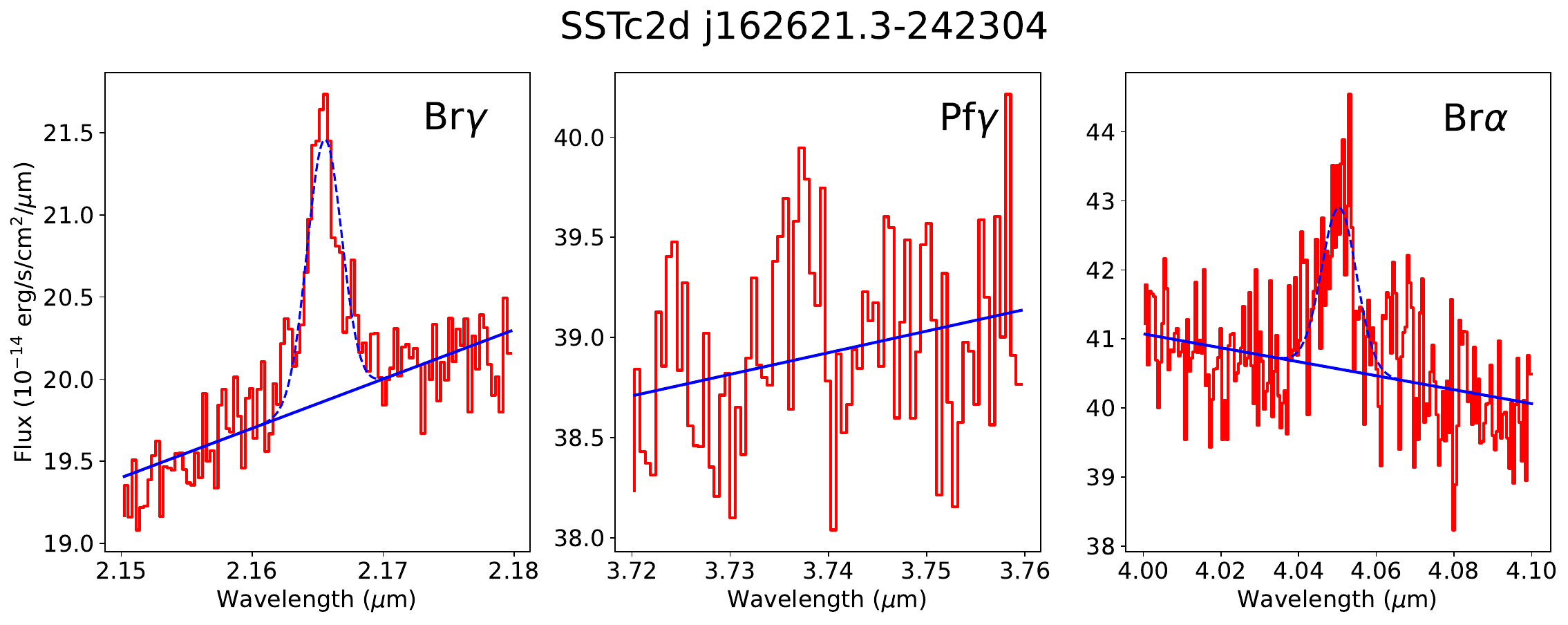} 
      \,\,\,\,
      \includegraphics[width=8.8cm]{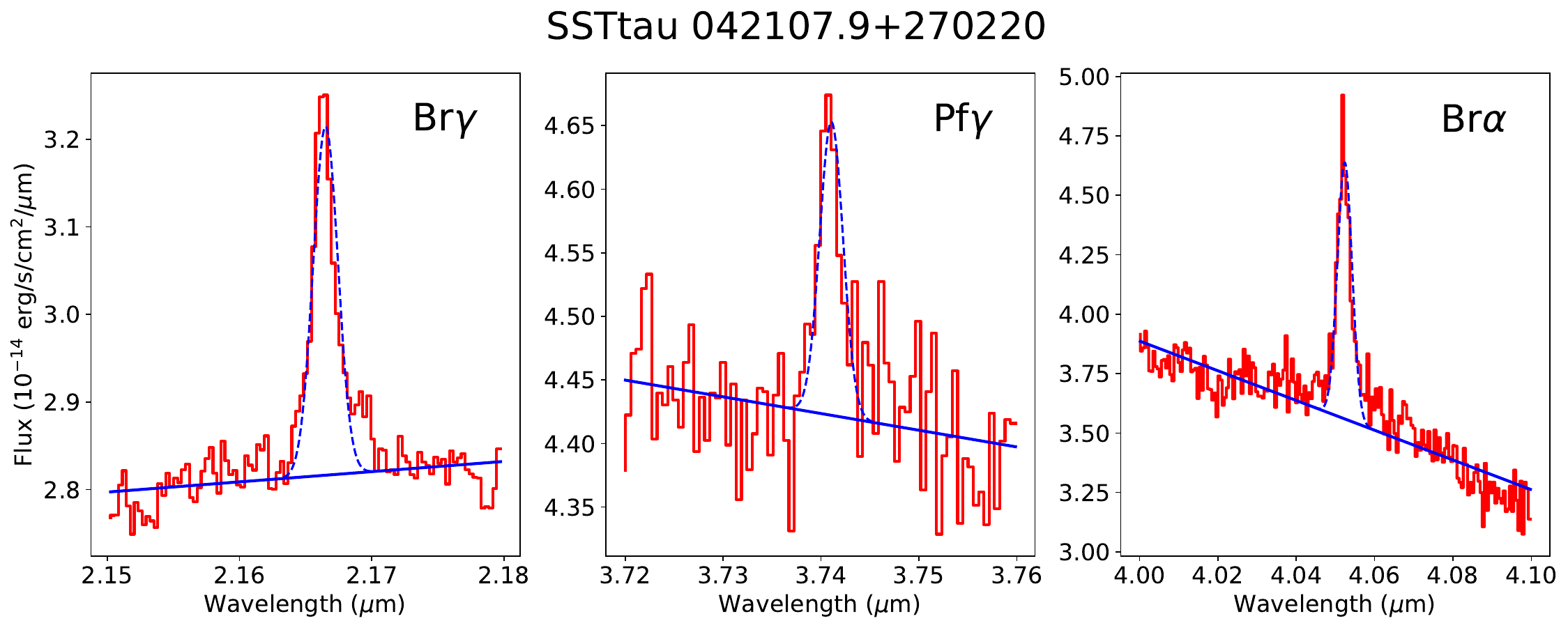} 
    }
    \centerline{
      \includegraphics[width=9.cm]{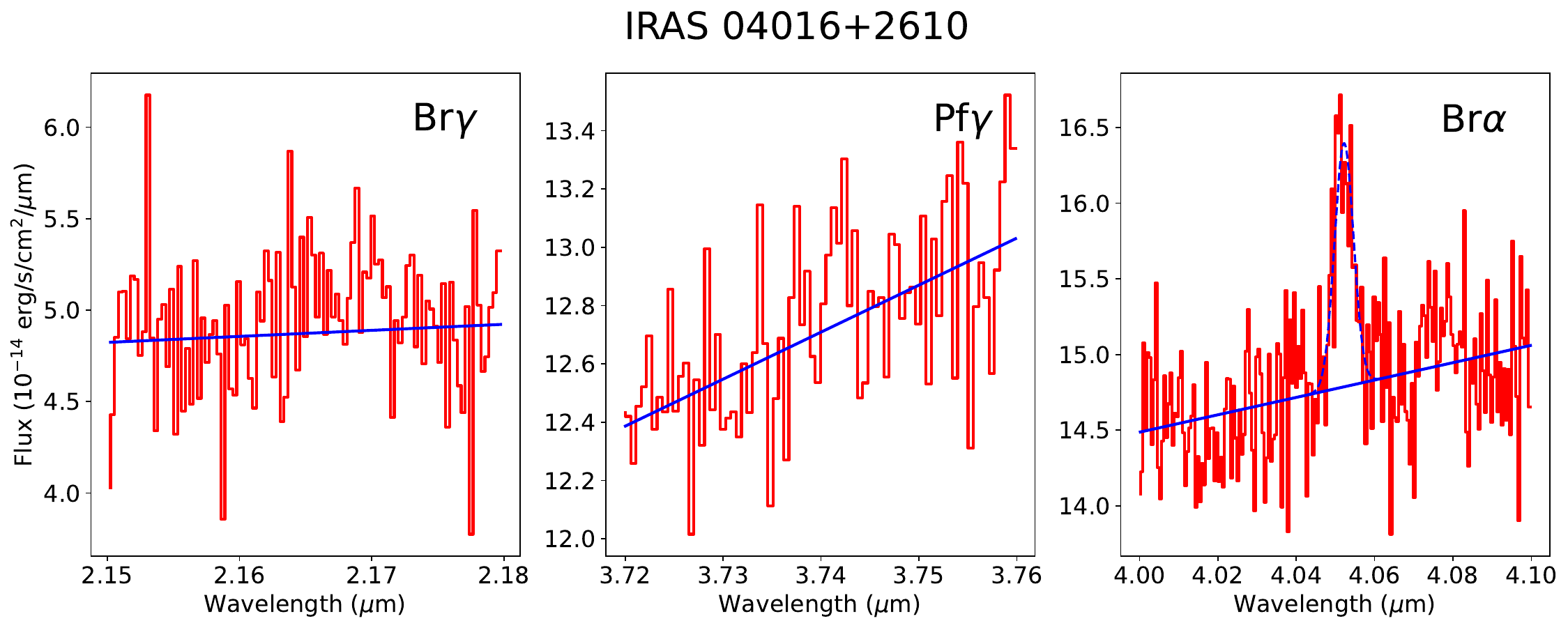} 
      \,\,\,\,
      \includegraphics[width=8.8cm]{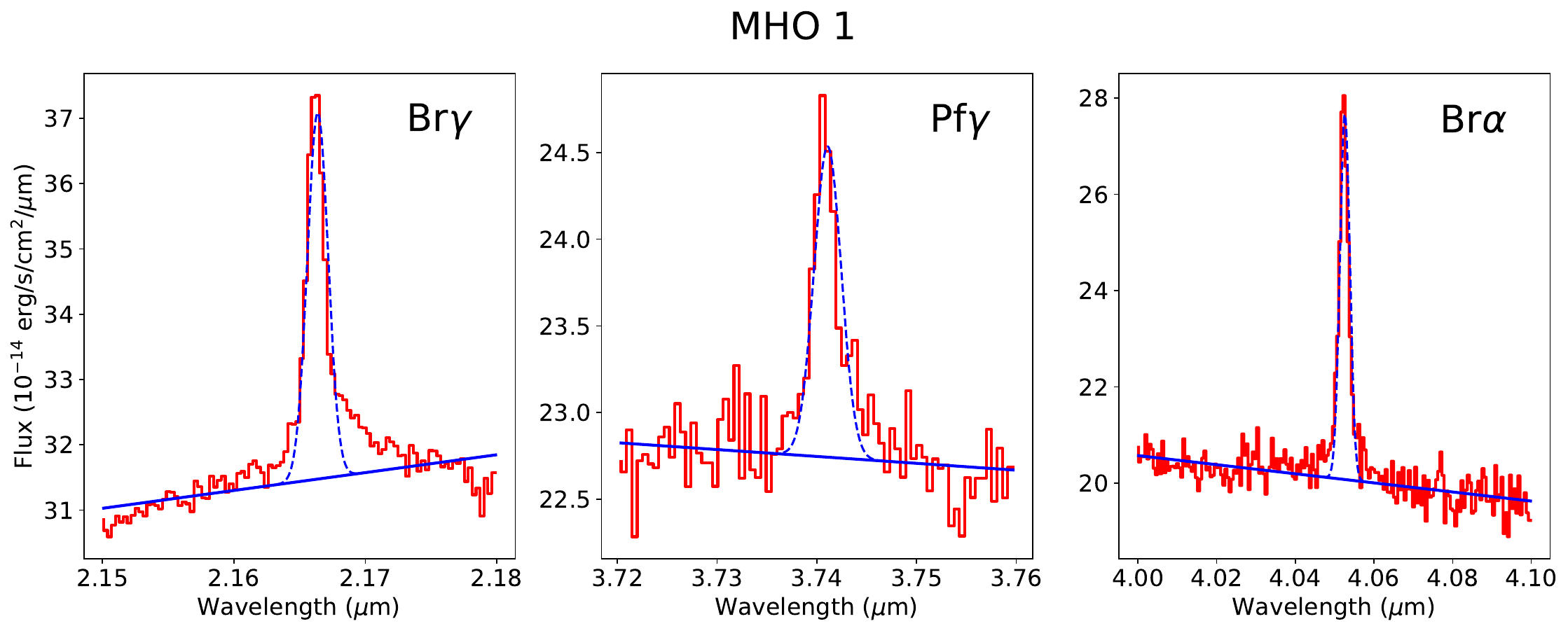} 
    }
    \caption{Same as Fig.~\ref{fig:f_class_ii_spectra} but for Class~I sources observed with SpeX.}
    \label{fig:f_class_i_spectra_a}

\end{figure*}

\begin{figure*}[!ht]
    \centerline{
      \includegraphics[width=8.8cm]{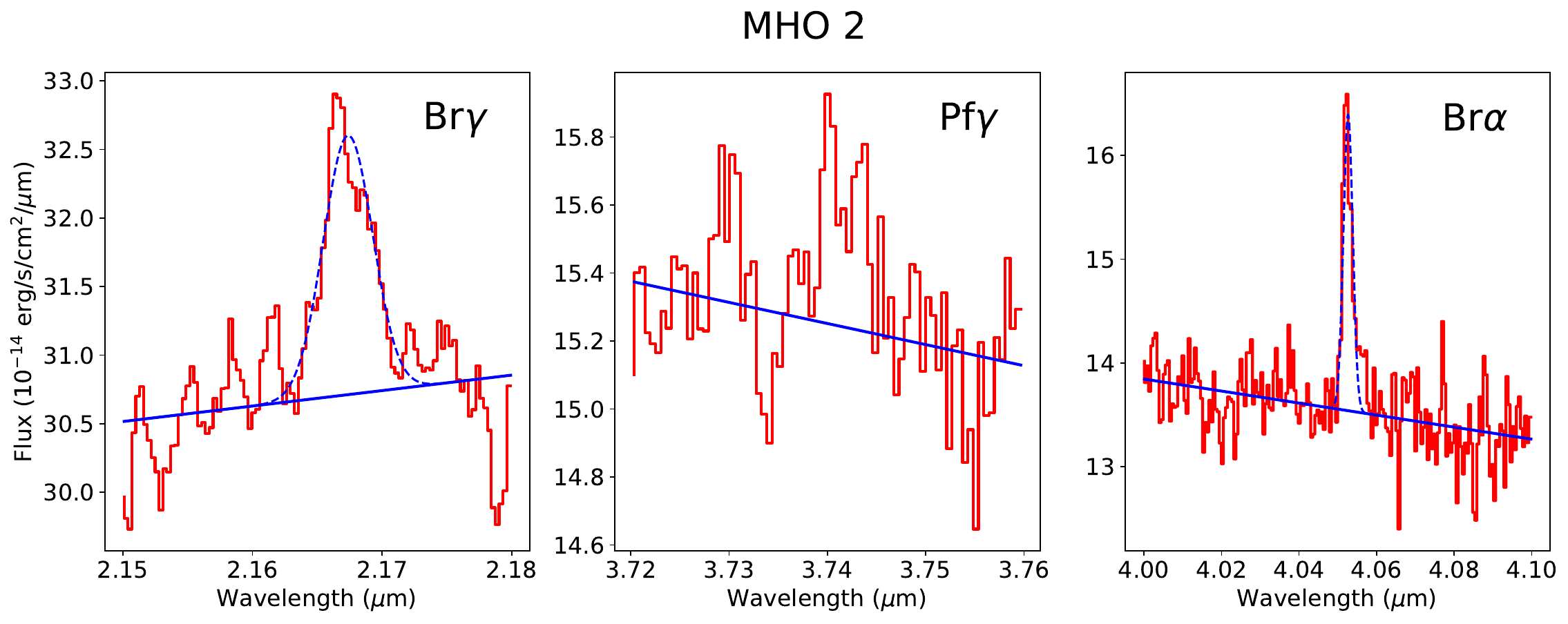} 
      \,\,\,\,
      \includegraphics[width=8.8cm]{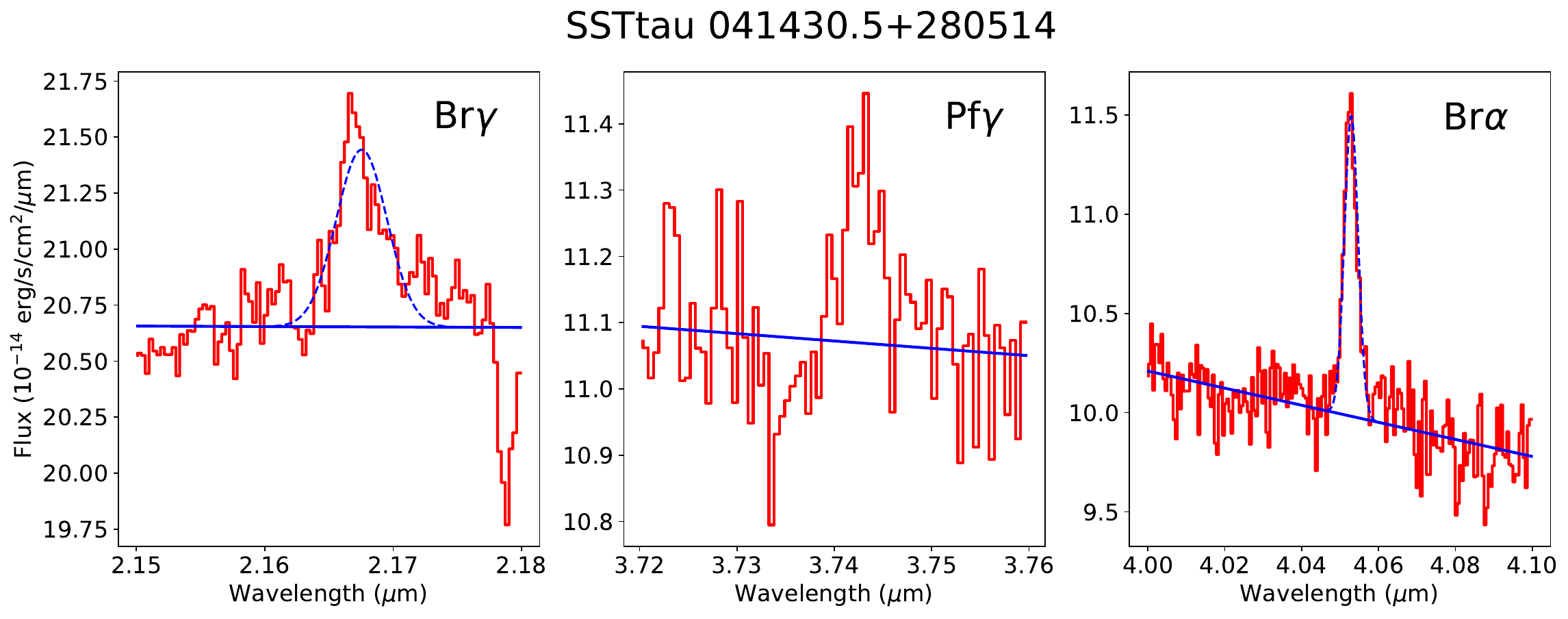} 
    }
    \centerline{
      \includegraphics[width=9.cm]{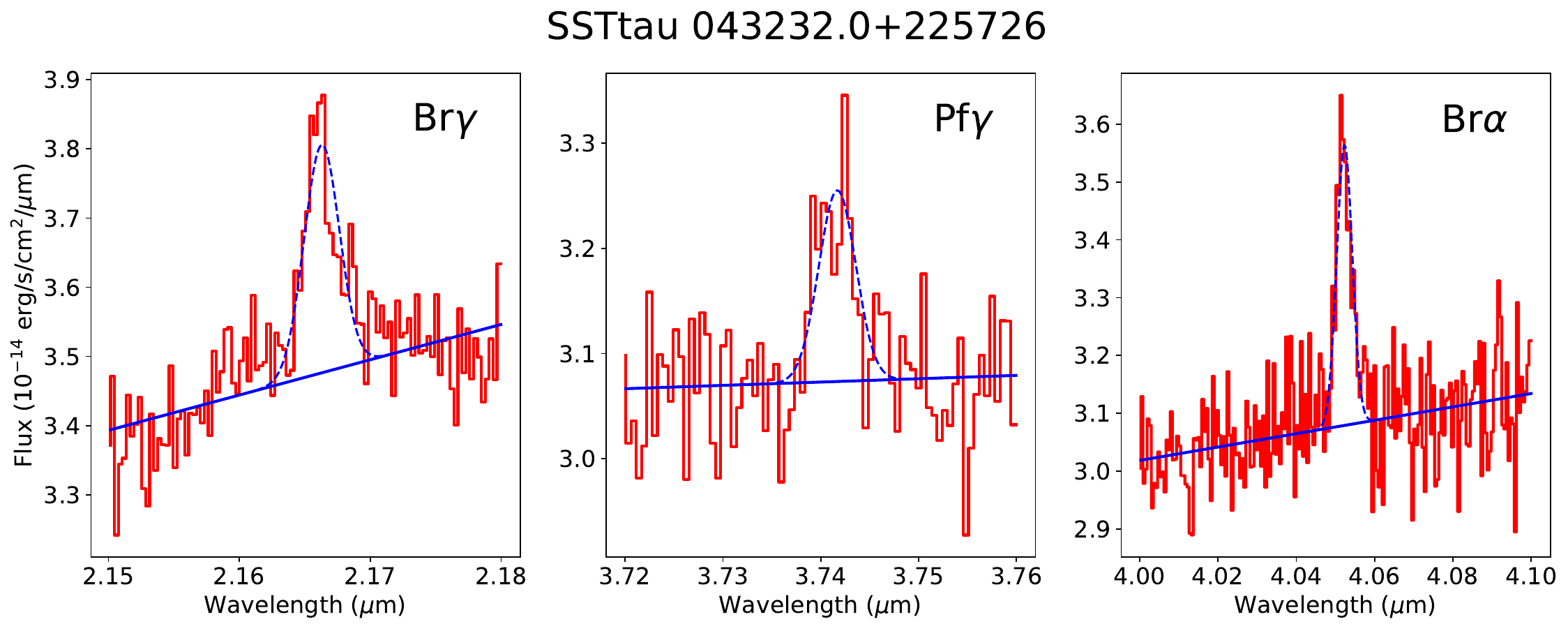} 
      \,\,\,\,
      \includegraphics[width=8.8cm]{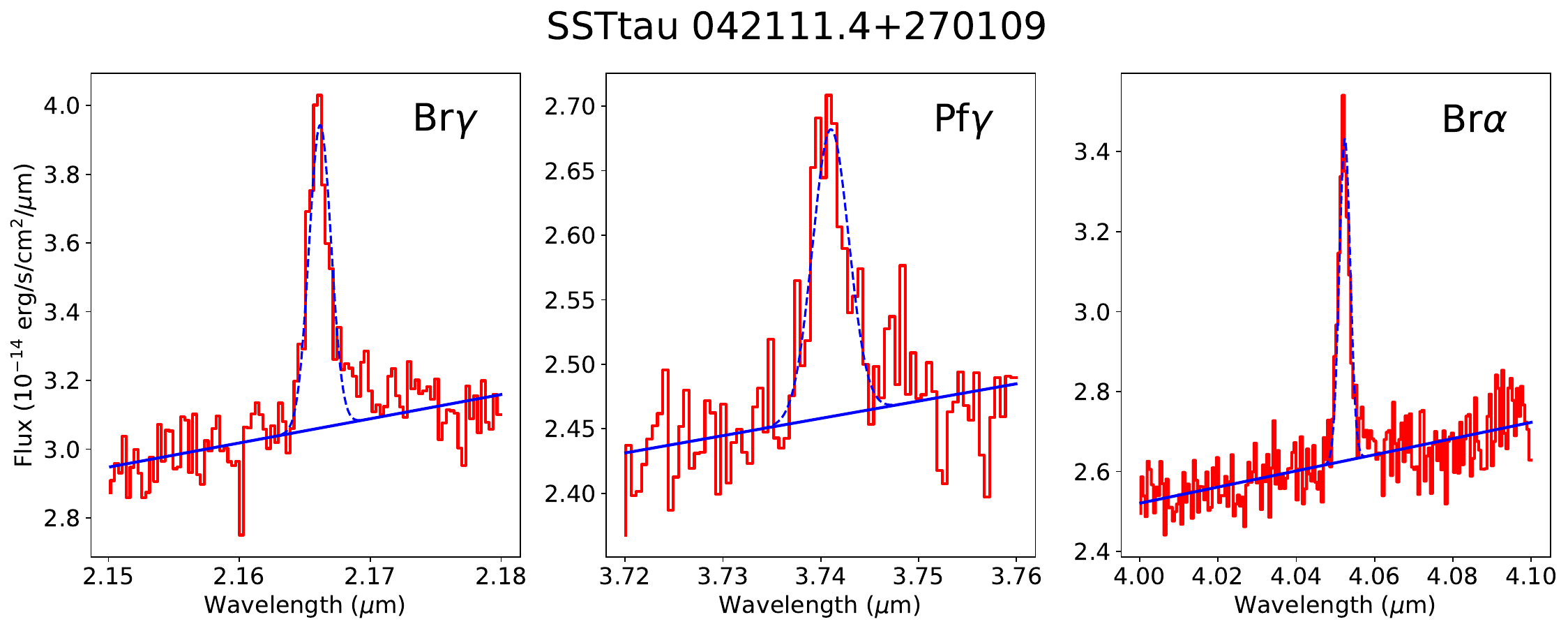} 
    }
      \includegraphics[width=9.cm]{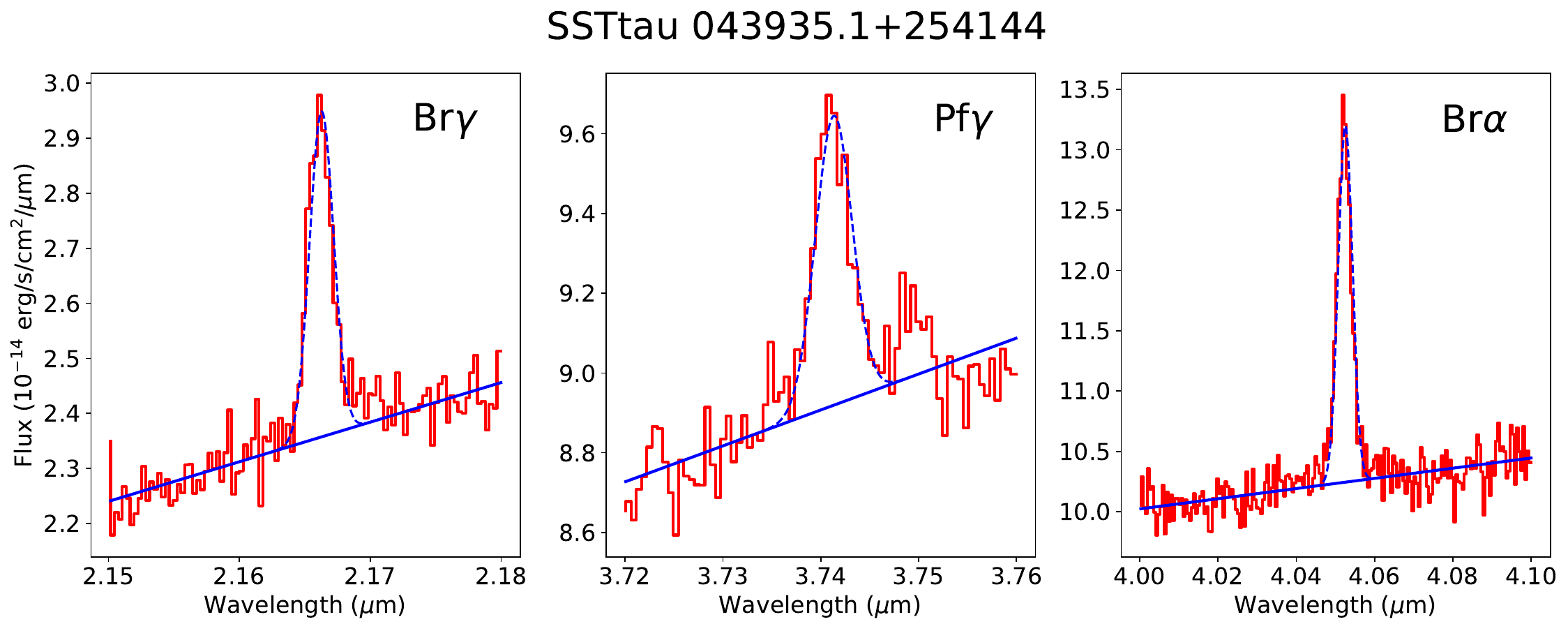} 
       

    \caption{Continued.}
    \label{fig:f_class_i_spectra_b}
\end{figure*}


\section {\Lacc-\LBrg\ correlation in Class~II}

\label{sec:app_lacc_lbrg}

As discussed in Sect.~\ref{Sec:ClassII_ratios}, several literature studies have reported correlations between the accretion rate luminosity, L$_{acc}$, and the intrinsic, extinction corrected Br$_\gamma$ line luminosity, \LBrg. The most complete and widely used studies are the one by \citet{2017A&A...600A..20A} for stars below about one solar mass, and the one by \citet{2017MNRAS.464.4721F} for intermediate mass stars. None of these two studies covers the full range of L$_{Br\gamma}$ measured in our sample of Class II, and since the two correlations presented in the literature show small, but significant differences we decided to recompute a correlation optimized for our range of measured line luminosities.

\begin{figure}[!ht]
    \includegraphics[width=8.8cm]{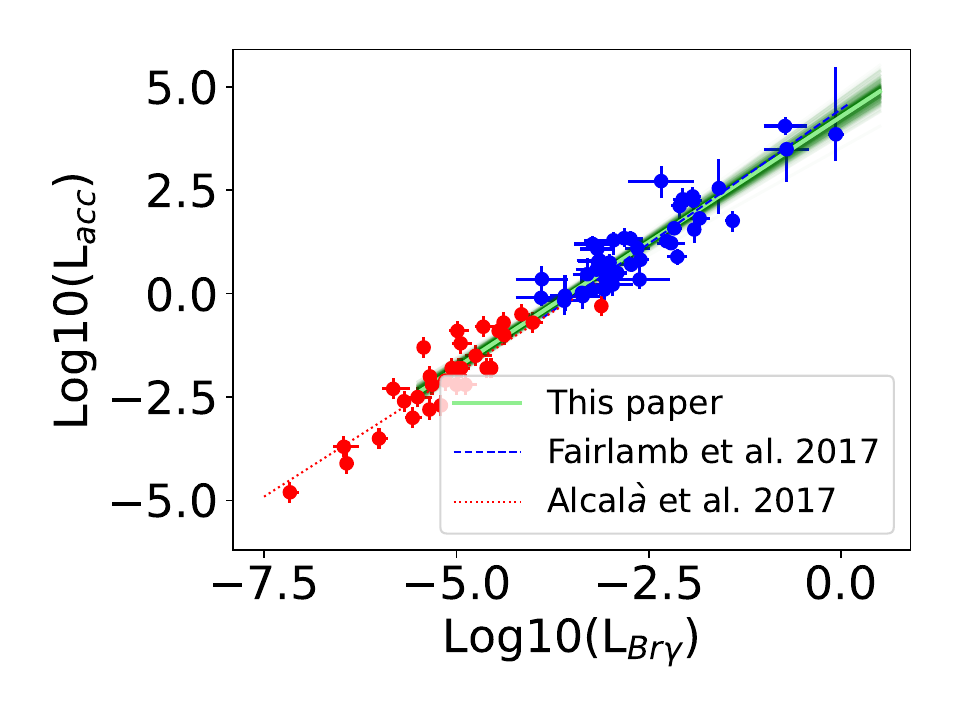} 
    \caption{Correlation of L$_{acc}$-L$_{Br\gamma}$ for Class~II objects in the literature. In red we show measurements for stars below about one solar mass \citep[from][]{2017A&A...600A..20A}, in blue for intermediate mass stars \citep[from][]{2017MNRAS.464.4721F}. The red dotted line and blue dashed lines show the literature correlations, derived for each sample separately. The light green line shows the correlation we derive for the range $-5.5\le ({\rm L}_{Br\gamma}/{\rm L}_\odot)\le0.5$. The dark green band shows the uncertainty in the fit.}
    \label{fig:f_lacc_lbrg_classii}
\end{figure}

In Figure~\ref{fig:f_lacc_lbrg_classii} we show the samples of  \Lacc\ and \LBrg\ measurements from \citet{2017A&A...600A..20A} and \citet{2017MNRAS.464.4721F}, the literature fits to the two separate samples, and our combined fit in the line luminosity range $-5.5\le$\LBrg$/{\rm L}_\odot\le0.5$. In deriving the fit, we only considered the good measurements reported in the literature, excluding upper limits in any of the two quantities. To perform the fit, we used the methodology described in \citet{2007ApJ...665.1489K}, using the Python implementation available at {\tt https://github.com/jmeyers314/linmix.git}.
The correlation, and its uncertainties, that we derived for use in this paper is
\begin{equation}
{\rm Log}_{10}({\rm L}_{\rm acc}/{\rm L}_\odot) = (1.21\pm 0.05)\, {\rm Log}_{10}({\rm L}_{Br_\gamma}/{\rm L}_\odot) \, + (4.28\pm 0.18).
\end{equation}

\end{appendix}

\end{document}